\documentclass[useAMS,usenatbib]{mnras}
\usepackage{times}

\newcommand{\msun}{\, \mathrm{M_{\odot}}}
\newcommand{\msunperyear}{\, \mathrm{M_{\odot}}\mathrm{yr}^{-1}}
\newcommand{\kms}{\, \mathrm{km}\mathrm{s}^{-1}}

\def\mnras{MNRAS}
\def\apj{ApJ}
\def\apjl{ApJ}

\def\aap{A\&Ap}

\def\aj{AJ}

\def\spose#1{\hbox to 0pt{#1\hss}}
\def\lta{\mathrel{\spose{\lower 3pt\hbox{$\mathchar"218$}}
     \raise 2.0pt\hbox{$\mathchar"13C$}}}
\def\gta{\mathrel{\spose{\lower 3pt\hbox{$\mathchar"218$}}
     \raise 2.0pt\hbox{$\mathchar"13E$}}}

\usepackage[dvips]{epsfig}
\usepackage{graphicx}
\usepackage{times}
\title[Gas Inflows due to Small Satellites]{Accretion of Small Satellites and Gas Inflows in a Disc Galaxy }

\author[Ram\'on-Fox \& Aceves]{F. G. Ram\'on-Fox$^{1,2,3}$ \thanks{E-mail: fgr2@st-andrews.ac.uk} and H\'ector Aceves $^1$.
\\
$^1$Instituto de Astronom\'{\i}a, Universidad Nacional Aut\'onoma de M\'exico, Apdo Postal 106,  Ensenada, Baja California, 22860 M\'exico \\
$^2$Scottish Universities Physics Alliance (SUPA), School of Physics and Astronomy, University of St Andrews, North Haugh, St. Andrews, Fife KY16 9SS, UK \\
$^3$ AIM, CEA, CNRS, Universit\'e Paris-Saclay, Universit\'e Paris Diderot, Sorbonne Paris Cit\'e, 91191 Gif-sur-Yvette, France}
\date{Accepted 1988 December 15. Received 1988 December 14; in original form 1988 October 11}

\pagerange{\pageref{firstpage}--\pageref{lastpage}} \pubyear{2002}

\begin{document}
%%%%%%%%%%%%%%%%%%%%%%%%%%%%%%%%%%%%%%%%%%%%%%%%%%%%%%%%%%%%%%%%%%%%%%%

\label{firstpage}

\maketitle

%%%%%%%%
\begin{abstract}
Galaxy interactions can have an important effect in a galaxy's evolution. Cosmological models predict a large number of small satellites around galaxies. It is important to study the effect that these small satellites can have on the host. The present work explores the effect of small $N$-body spherical satellites with total mass ratios in the range $\approx$ 1:1000-1:100 in inducing gas flows to the central regions of a disc galaxy with late-type morphology resembling the Milky Way. Two model galaxies are considered: barred and non-barred models; the latter one is motivated in order to isolate and understand better the effects of the satellite. Several circular and non-circular orbits are explored, considering both prograde and retrogade orientations. We show that satellites with such small mass ratios can still produce observable distortions in the gas and stellar components of the galaxy. In terms of gas flows, the prograde circular orbits are more favourable for producing gas flows, where in some cases up to $60\%$ of the gas of the galaxy is driven to the central region. We find, hence, that small satellites can induce significant gas flows to the central regions of a disc galaxy, which is relevant in the context of fuelling active galactic nuclei.
\end{abstract}

%%%%%%%%%%%%%%%%%%%%%%%%%%%%%%%%%%%%%%%%%%%%%%%%%%%%%%%%%%%%%%%%%%%%%%%%%
%%%%%%%%%%%%%%%%%%%%%%%%%%%%%%%%%%%%%%%%%%%%%%%%%%%%%%%%%%%%%%%%%%%%%%%%%

\large

%%%%%%%%%%%%%%%%%%%%%%%%%%%%%%%%%%%%%%%%%%%%%%%%%%
%%%%%%%%%%%%%%%%%%%%%%%%%%%%%%%%%%%%%%%%%%%%%%%%%%
%%%%%%%%%%%%%%%%%%%%%%%%%%%%%%%%%%%%%%%%%%%%%%%%%%
\section{Introduction}\label{sec:intro}

Galaxies are not isolated from the environment where they reside and constitute an open system where different kind of physical interactions occur; in particular, gravitational interactions with other galaxies, both major or minor, are expected to occur during their whole lifespan. Actually, it is rather hard to define in practice what an isolated galaxy is and where to study truly secular phenomena (e.~g.~\citealt{Hopkinsetal2010}). These interactions can play an important role in long term galaxy evolution.

The study of the effects of major gravitational interactions on the structure and evolution of galaxies has had a long tradition. Initial simulations began considering only collisionless $N$-body simulations (e.~g.~\citealt{Holmberg1941} and \citealt{ToomreandToomre1972}). The modelling of galaxy interactions has been increasing in complexity both in including the physics of the interaction as well as the numerical methods. Some works began exploring the effects of gas dynamics in the evolution of mergers (e.~g.~\citealt{BarnesandHernquist1991}, \citealt{BarnesandHernquist1996}, \citealt{Barnes2002}). The increase in computational power has allowed to include other physical effects such as star formation and feedback (e.~g.~\citealt{Springel2000}, \citealt{Coxetal2004}, \citealt{SpringelandHernquist2005}), as well as metallicity effects (e.~g.~\citealt{Torreyetal2012,Bustamante_etal2018}). Some works have included the effects of magnetic fields in a three-galaxy interaction (e.~g.~\citealt{Kotarbaetal2011}). Other works have compared the differences between different hydrodynamical approaches (e.~g.~\citealt{DiMatteoetal2008, Gaboretal2016}). Recent works have started to explore the evolution of major mergers with sub-parsec resolution (e.~g.~\citealt{Renaudetal2015}). These are just some examples as this has been a widely studied topic.

Many galaxies present an Active Galactic Nucleus (AGN) (e.~g.~\citealt{AlexanderHickox2012}), which require large amounts of gas to be driven down to the inner parsecs of a galaxy (e.~g.~\citealt{Shlosmanetal90}). Several observational works have found that a fraction of galaxies hosting an AGN also have nearby companions (e.~g.~\citealt{Satyapaletal2014,Gouldingetal2018}), suggesting that interactions may play a role in driving inward gas flows. Some numerical simulations have also shown, for example, that major interactions between galaxies with a gas component are able to induce large torques on the gas component and transport large amounts of angular momentum to the centres of galaxies, with the ability to fuel an AGN (e.~g~\citealt{BarnesandHernquist1996,BlumenthalandBarnes2018}).

Minor mergers may be an AGN-triggering mechanism (e~g.~\citealt{Taniguchi1999}). In addition, some observations show that a fraction of galaxies with disturbed morphologies also have an enhanced star formation rate (SFR). For example, \citet{Rudnicketal2000} show that there is some correlation between lopsidedness and an increased SFR; a similar trend is seen in galaxies with disturbed spiral arms (e.~g.~\citealt{Edmanetal2012,Kaviraj2014}).

The previous examples motivate a deeper analysis of minor mergers, but the study of these interactions has progressed at a somewhat slower rate. Several works have shown that they can lead to detectable and important morphological features in the evolution of spiral galaxies such as lopsidedness (e.~g.~\citealt{Bournaudetal05}), grand-design spiral arms (e.~g.~\citealt{Dobbsetal2010,Pettittetal2016,Pettittetal2018}) as well as disturbed spiral arms (e.~g.~\citealt{Starkenburgetal2016}), ring-like features in S0 galaxies (e.~g.~\citealt{Mapellietal2015}), as well as hole features in the HI distribution (e.~g.~\citealt{BekkiChiba2006,Kannanetal2012,Shahetal2019}). \citet{Coxetal2008} showed that the triggered starburst decreases when the satellite's mass is reduced. From the Galaxy's point of view, they can also be a mechanism that forms stellar streams and substructures in the halo (e.~g.~\citealt{Purcelletal2011,Loraetal2012}). The previous studies have explored effects of satellites with total mass ratio to the main galaxy (${\cal R}$) in the range of $1:4$ to about $1:1000$.

Within the cosmological context, the mass function of satellites around galaxies is dominated in number by a large quantity of small faint satellites (e.~g.~\citealt{Klypinetal1999,Mooreetal1999,Gonzalezetal2006,WangandWhite2012,Sawalaetal2017}). So a question that naturally arises in this context is: down to what mass ratios are the effects of small satellites still noticeable and significant in the evolution of the host galaxy? Particularly, what effects can very small satellites in the range of total mass ratios of ${\cal R} \approx$~$1:100$ have in the large-scale motions of gas? This last question bears relation to the possibility of moving important amounts of gas to the inner parts of disc galaxies to provide a condition for further movement into the central parts by local mechanisms. 

Works exploring the above questions are the classical studies of \citet{MihosHernquist94} and \citet{HernquistMihos95}. They investigated the effects of the impact of small satellites, with ${\cal R}\approx 1\!:\!50$, on the large-scale gas motions under different orbits. They found that prograde encounters in circular orbits lead to integrated flows of $\approx 10^{9} {\rm M}_\odot$ in about a few orbital periods. Other orbits may be less effective in moving such gas quantities to the inner regions. Recent works have focused on reproducing features in spiral galaxies expected to be produced by minor interactions (e.~g.~\citealt{Dobbsetal2010,ChakrabartiandBlitz2009,Chakrabarti11,Pettittetal2016,HuandSijacki2018,Shahetal2019}).

To our knowledge, the exploration of the effects of very small ratios (${\cal R}< 1:100$) have in triggering large-scale flows in spiral galaxies has only been explored in a few works up to now, and it can still be addressed in much more detail. It is probable that during a disc galaxy's cosmic history, it suffers multiple perturbations by this type of satellites, and their individual or collective effects may play a role in the large-scale gas dynamics and other effects such as star formation.

In this work, we explore the gas flows in a disc galaxy induced by the perturbation of a satellite with a mass ratio ${\cal R}$ between 1:1000--1:100. In \S \ref{sec:model}, we describe the models, initial conditions, and simulation parameters. In \S \ref{sec:results}, the results of the simulations are described. A discussion of these is presented in \S \ref{sec:discussion} and \S \ref{sec:conclusions} summarizes the conclusions.

%%%%%%%%%%%%%%%%%%%%%%%%%%%%%%%%%%%%%%%%%%%%%%%%%%
%%%%%%%%%%%%%%%%%%%%%%%%%%%%%%%%%%%%%%%%%%%%%%%%%%
%%%%%%%%%%%%%%%%%%%%%%%%%%%%%%%%%%%%%%%%%%%%%%%%%%
\section{Models and Initial Conditions}
\label{sec:model}

%%%%%%%%%%%%%%%%%%%%%%%%%%%%%%%%%%%%%%%%%%%%%%%%%%
%%%%%%%%%%%%%%%%%%%%%%%%%%%%%%%%%%%%%%%%%%%%%%%%%%
\subsection{Galaxy Models}
\label{subsec:galmodels}

%%%%%%%%%%%%%%%%%%%%%%%%%%%%%%%%%%%%%%%%%%%%%%%%%%
\subsubsection{Primary Galaxy}

The method and code of \citet{McMillanDehnen07} is used to generate isolated disc galaxies or spheroidal systems.\footnote{This code is available through the NEMO Stellar Dynamics Toolbox \citep{Teuben95}} The primary galaxy consists of a self-consistent dark matter halo, a stellar disc, and a bulge. The dark matter halo is represented by a truncated \citet{NFW96} (NFW) density profile given by:
\begin{equation}
	\rho_h(r) = \frac{\rho_0}{(r/r_h) (1 + r/r_h)^2} \mathrm{sech}(r/r_t)
\end{equation}
where $r_h$ is the scale radius and $r_t$ defines the scale length of the truncation function. The disc has an exponential-isothermal sheet density profile given by:
\begin{equation} 
	\rho_d(R,z) = \frac{M_d}{4 \pi R_d^2 z_d} \exp\left(-\frac{R}{R_d} \right) \mathrm{sech}^2 \left( \frac{z}{z_d} \right)
\end{equation}
where $M_d$ is the disc mass, $R_d$ is the scale radius, and $z_d$ is the vertical scale. The stellar bulge follows a \citet{Hernquist90} profile given by:
\begin{equation}
	\rho_b(r) = \frac{M_b }{2 \pi r_b^3} \frac{1}{(r/r_b) (1 + r/r_b)^3}
\end{equation}
where $M_b$ is the mass of the bulge and $r_b$ is the scale radius. 

For the halo, we set $M_h = 10^{12} \msun$, $r_h = 21$ kpc, and $r_t = 210$ kpc. The disc has a mass of $M_d = 4.167 \times 10^{10} \msun$, a scale radius of $R_d = 3.5$ kpc, and a vertical scale of $z_d = 0.35$ kpc. The mass of the bulge is $M_b = 8.33 \times 10^{9} \msun$ and the scale radius is $r_b = 0.7$kpc. These parameters are representative for a Milky Way sized galaxy \citep{KlypinZhaoSomerville02,McMillanDehnen07}. 

We also include an isothermal gas disc with $M_g = 0.1 M_d = 4.167 \times 10^9 \msun$ and $T = 10^4$ K. It is initially distributed in a similar way to the stellar disc. The circular velocity of the gas has been initialised taking into account the effect of a radial pressure gradient due to the initial radial density profile (e.~g. \citealt{Wang10})

The method of \citet{McMillanDehnen07} allows to set the Toomre $Q$ parameter to be a nearly constant function of radius. We choose two models, one with $Q = 3.0$ (Model~A) and a second one with $Q = 1.5$ (Model~B). The first model, although rather unrealistic for a spiral galaxy like the Milky Way, is advantageous for isolating as much as possible the effects of the perturbing satellite from those of galactic structure such as spiral arms and a central bar. Model~B forms a bar which drives gas flows in the inner galaxy in isolated evolution.

%%%%%%%%%%%%%%%%%%%%%%%%%%%%%%%%%%%%%%%%%%%%%%%%%%
\subsubsection{Satellite Galaxy}

The satellite galaxy is assumed to be a pure dark matter subhalo represented by a Plummer profile:
\begin{equation}
	\rho_s(r) = \frac{\rho_0}{(1 + (r/a)^2)^{5/2}}
\end{equation}
where $a$ is the scale radius, $\rho_0 = 4 M_s/4 \pi a^3$, and $M_s$ is the mass of the satellite. The density and scale radius of the satellite are defined such that the average density within its half-mass radius is similar to that of the disc. A similar approach has been followed by \citet{MihosHernquist94} and \citet{HernquistMihos95} in specifying the parameters of a satellite. This has the advantage of using satellites with different masses while preserving a constant-average density. The following set of parameters are chosen for the model satellite:  $M_s = 6 \times 10^9 \msun$ and $a = 1.0$ kpc, labelled Satellite 1, and $M_s = 1.2 \times 10^{10} \msun$ and $a = 1.3$ kpc, labelled Satellite 2. These are chosen in order to test the effect of varying the mass of the satellite on the induced flows in the disc. These masses correspond to mass ratios with respect to the total mass of the galaxy of ${\cal R}\approx 6:1000$ and $\approx 3:265$ respectively.

The orbital parameters for the infalling satellites are chosen as follows. First, we take a circular orbit inclined by $30^{\circ}$ and test two initial orbital radii: $R_i = 3R_d$ and $R_i = 6R_d$ (labelled Orbit 1 and 2, respectively). Prograde and retrograde encounters are explored for both orbits. The prograde encounter with $i = 30^{\circ}$ and $R_i = 6R_d$ has also been explored in \citet{HernquistMihos95}. This choice of orbits allows to explore the effect of the impact parameter of the satellite in inducing gas flows in the host. In order to explore the effects of non-circular and coplanar orbits, we assume an orbit with apocenter $r_a \approx 6 R_d$ kpc and pericenter $r_p = R_d$ (Orbit 3). Prograde and retrograde encounters are also  considered. We consider that this choice of orbits allows to test the effect of varying the impact of the satellite and compare the difference between an encounter in the central region to one in the outer region of the host. Although high-eccentricity orbits may be more likely for infalling satellites, low-eccentricity orbits are not discarded according to eccentricity distributions derived from cosmological simulations (e.g. \citealt{Benson2005,Zentneretal2005,KhochfarBurkert2006,Wetzel2011,Barberetal2014}).

%%%%%%%%%%%%%%%%%%%%%%%%%%%%%%%%%%%%%%%%%%%%%%%%%%
\subsection{Numerical Code and Simulation Parameters}

The simulations are performed with the public version of the $N$-body and hydrodynamics code GADGET-2 \citep{Springel05}. It is a highly parallelised code that implements the Tree code \citep{BarnesHut86} to compute gravitational forces. Gas dynamics are treated with the Smoothed Particle Hydrodynamics (SPH) (e.~g.~\citealt{GingoldMonaghan77,Price2012}) method and gravitational forces are computed using an optimised version of the TreeSPH code \citep{HernquistKatz1989}. This version of the code allows the user to implement either an isothermal or an adiabatic equation of state for an ideal gas to treat the gas thermal physics.

The models described in \S \ref{subsec:galmodels} are constructed with the following parameters. The components of the host galaxy were generated using $N_h = 10^6$ particles for the halo, $N_d = N_g = 500\,000$ for the stellar and gaseous disc respectively, and $N_b = 10\,000$ for the bulge. All satellite galaxy models are initialised with $N_s = 72000$ particles. This level of resolution is chosen since we are interested in evolving the system for long-time scales. All the simulations include gas self-gravity.

GADGET-2 allows to specify different gravitational softenings to particles of different components of the galaxy. The softening lengths used in the simulations are: $\epsilon_h = 0.121$ kpc for halo particles, $\epsilon_g = \epsilon_s = 0.035$ kpc for gas and stellar particles, and $\epsilon_b = 0.1$ kpc for bulge particles. The softening parameter for the satellite's particles has been set equal to that of the disc. The simulations are performed in a natural system of units $(G = 1)$ where the unit of mass is $u_m = 10^{10} \msun$, the unit of distance is $u_l = 1.0$ kpc, the unit of time is $u_t = 4.7$ Myr, and the unit velocity is $u_v = 207.5 \kms$. 

Both models A and B are evolved in isolation up to $t = 8\tau$, where $\tau \approx 240$ Myr is the orbital period of the model galaxy at $R = 8$ kpc in order to test the stability of the system. The positions and velocities at $t = 4\tau$ of this simulation are used as initial conditions for the host galaxy in the simulations that include the infalling satellite. This choice is motivated by the fact that for the barred model, the formation of the bar and growth of spiral structure takes some time. This is to ensure that the initial condition that includes the satellite has a morphology representative of a late-type barred galaxy. This satellite-galaxy system is evolved for an additional lapse of $4\tau$. 

%%%%%%%%%%%%%%%%%%%%%%%%%%%%%%%%%%%%%%%%%%%%%%%%%%
%%%%%%%%%%%%%%%%%%%%%%%%%%%%%%%%%%%%%%%%%%%%%%%%%%
%%%%%%%%%%%%%%%%%%%%%%%%%%%%%%%%%%%%%%%%%%%%%%%%%%
\section{Results}
\label{sec:results}

%%%%%%%%%%%%%%%%%%%%%%%%%%%%%%%%%%%%%%%%%%%%%%%%%%
%%%%%%%%%%%%%%%%%%%%%%%%%%%%%%%%%%%%%%%%%%%%%%%%%%
\subsection{Morphology}
\label{subsec:morphology}

% Prograde Orbit 2, M = 6e9 Msun
\begin{figure*}
	\centering
	\includegraphics[clip, trim=0 18pt 0 0, width=0.65\textwidth]{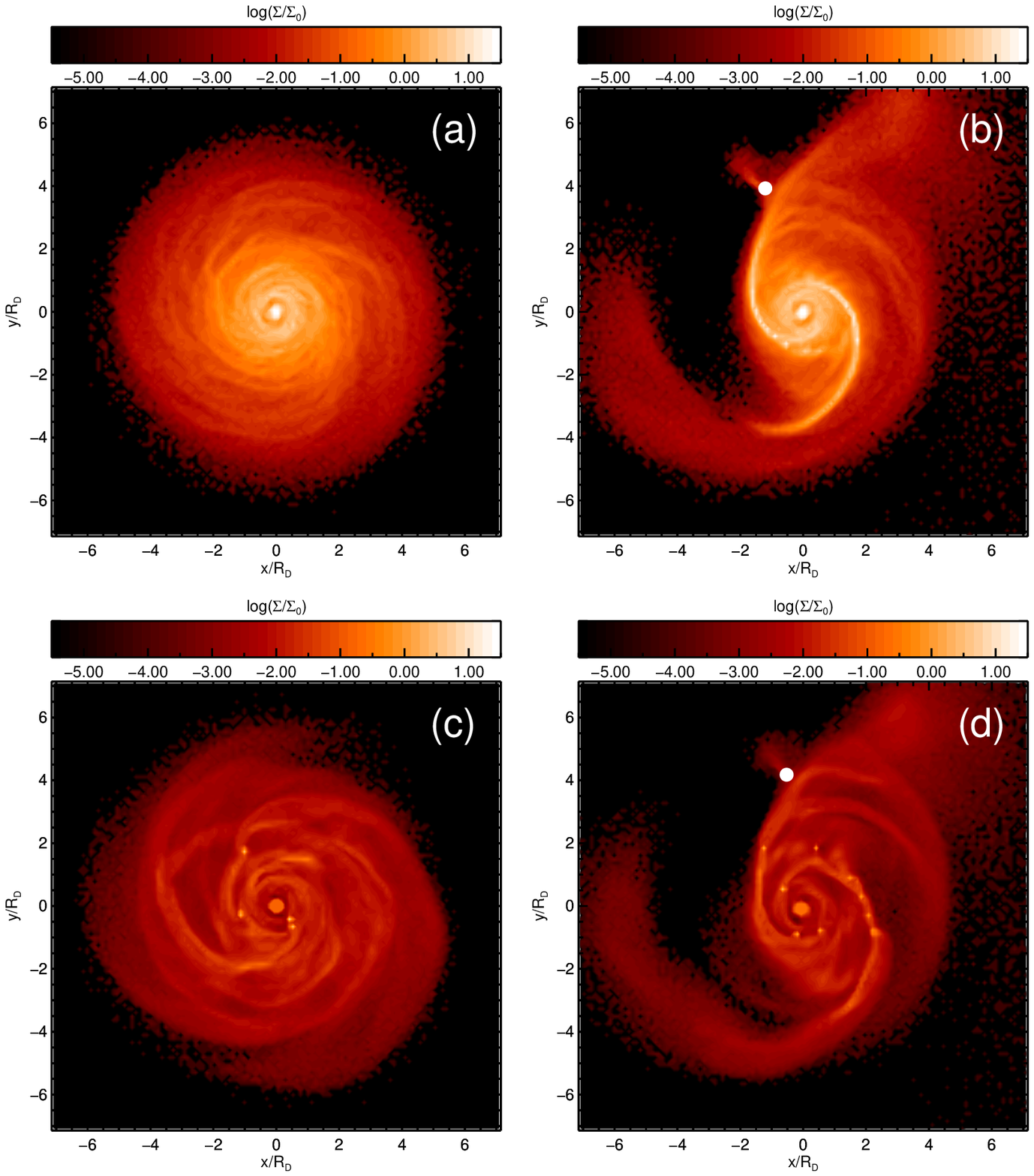}
	
	\includegraphics[clip, trim=0 18pt 0 0, width=0.65\textwidth]{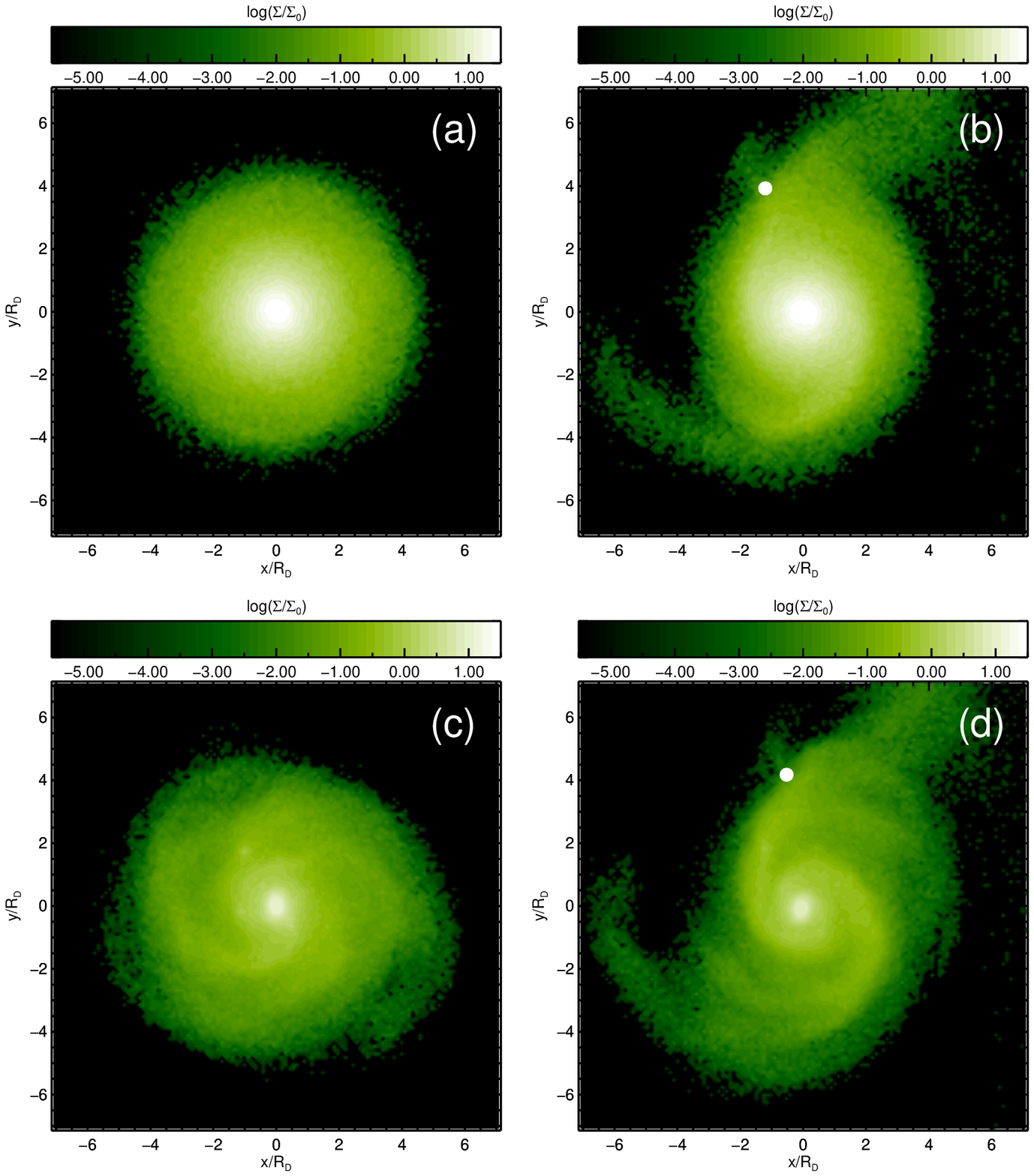}
	\caption{Morphology of the gaseous (top four panels, red map) and stellar (bottom four panels, green map) components produced with the $M = 6 \times 10^9 \msun$ satellite in a prograde encounter following Orbit 2. The maps show the logarithm of the surface mass density, $\Sigma$, normalized to its central value. Panels (a) and (b) show the isolated and interacting simulations with Model A, and panels (c) and (d) show the isolated and interacting simulations with Model B, respectively.}
	\label{fig:morph-orbit2-pro}
\end{figure*}

% Retrograde Orbit 1, M = 6e9 Msun
\begin{figure*}
	\centering
	\includegraphics[clip, trim=0 18pt 0 0, width=0.65\textwidth]{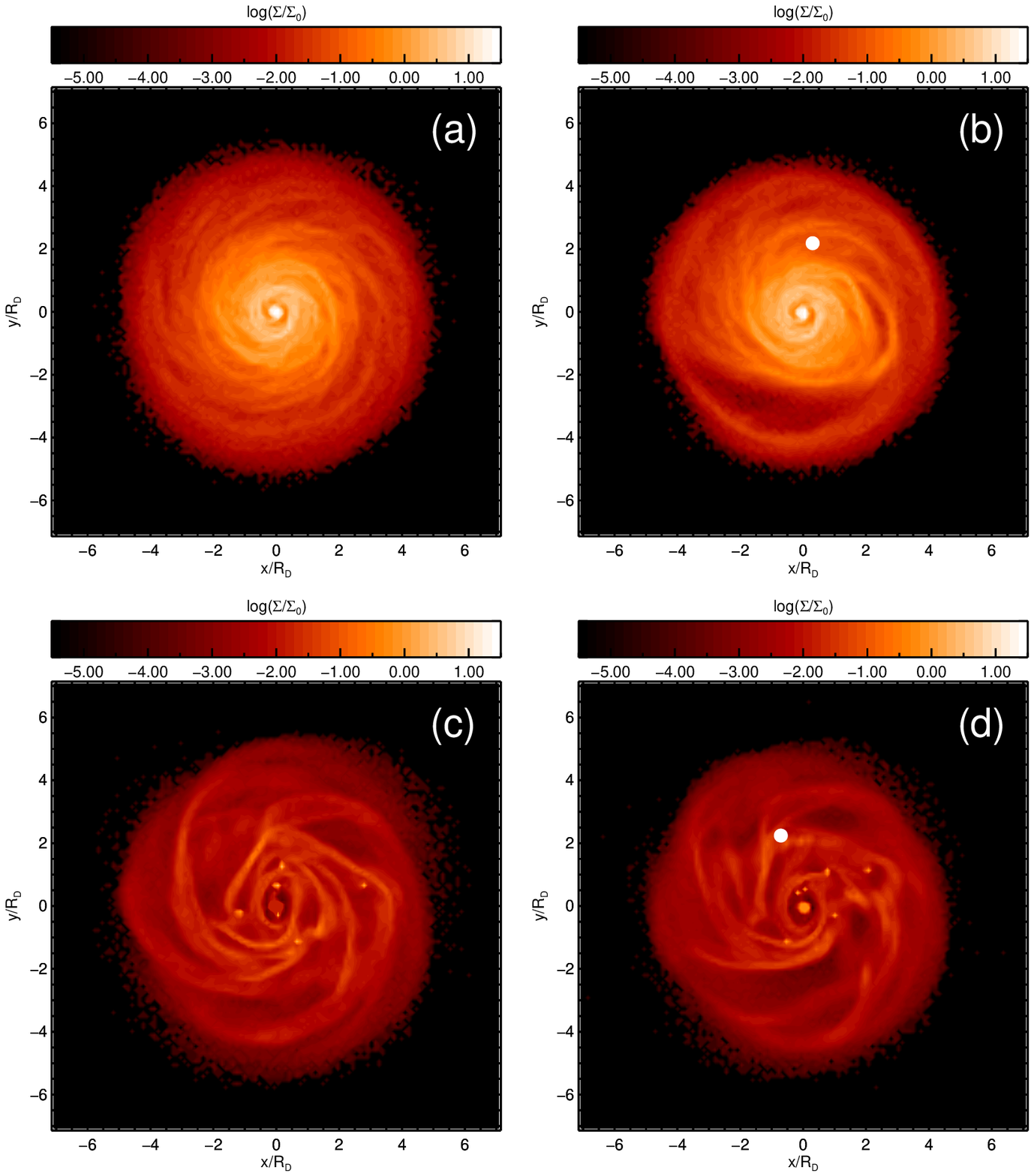}

	\includegraphics[clip, trim=0 18pt 0 0, width=0.65\textwidth]{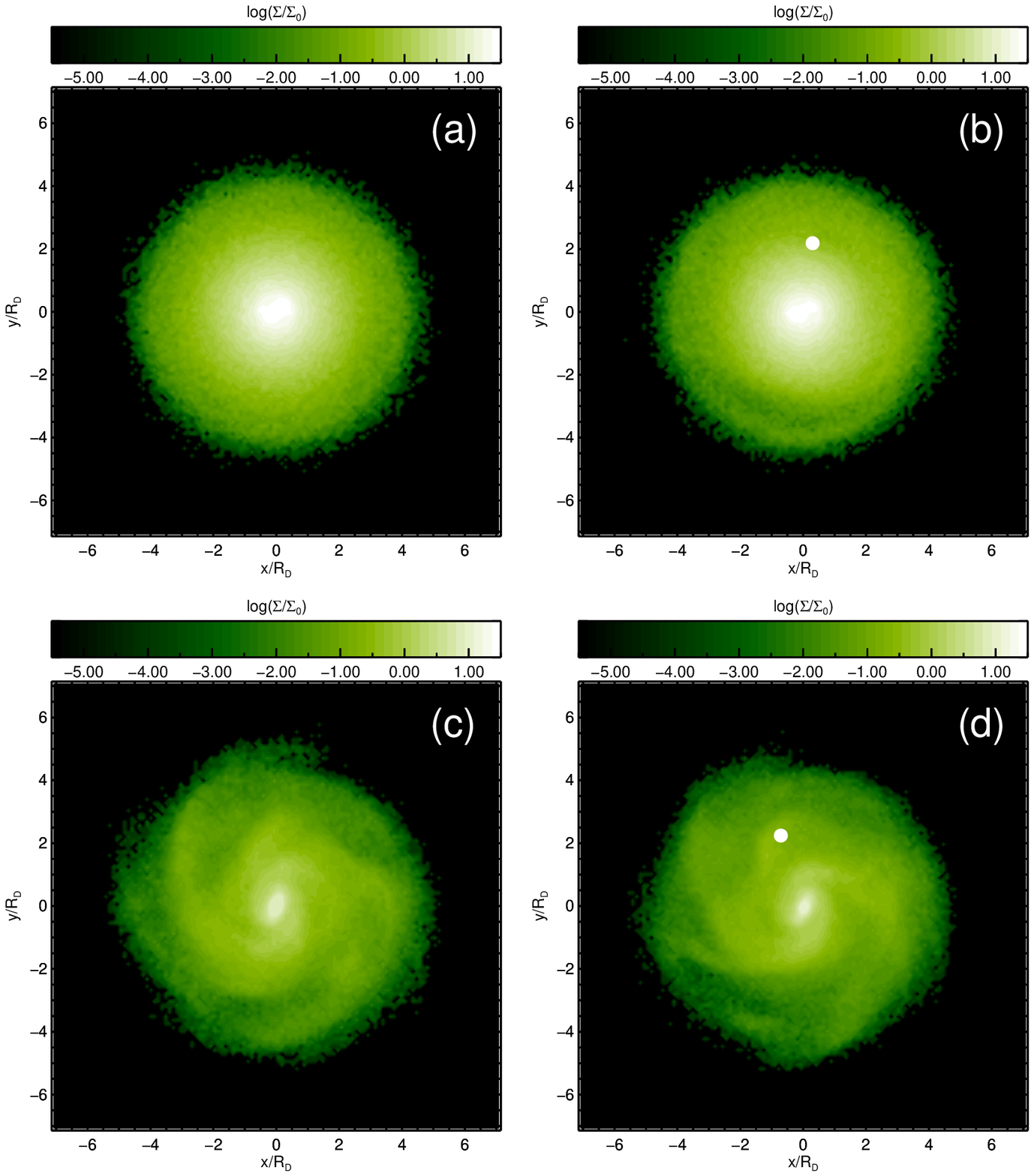}
	\caption{Morphology of the gaseous (top four panels, red map) and stellar (bottom four panels, green map) components produced with the $M = 6 \times 10^9 \msun$ satellite in a retrograde encounter following Orbit 1. The maps show the logarithm of the surface mass density, $\Sigma$, normalized to its central value. Panels (a) and (b) show the isolated and interacting simulations with Model A, and panels (c) and (d) show the isolated and interacting simulations with Model B, respectively.}
	\label{fig:morph-orbit1-ret}
\end{figure*}

The simulations described in \S \ref{sec:model} show that prograde encounters trigger the formation of spiral patterns on the host galaxy, with the most prominent induced by the satellite with the initial orbit with $R_i = 6R_D$ (Orbit 2).  As shown in the top right panels (b and d) of Figures~\ref{fig:morph-orbit2-pro} (gas) and the bottom right (b and d) \ref{fig:morph-orbit2-pro} (stars), it resembles the morphology of a grand design spiral galaxy.

For the non-barred galaxy (model A), the satellite in the smaller orbit produces morphological disturbances, though not as symmetric as those by the satellite in the larger orbit (Orbit 2). In the case of Orbit 2, some features are generated in the first disc crossings, and a strong $m = 2$ spiral pattern is produced after the third passage at $t \approx 2.9\tau$. The spiral pattern is still present after a fourth passage of the satellite. The latter eventually merges with the central region. In the case of the retrograde encounters with model A, the smaller orbit ($R_i = 3R_d$) produces some disturbances or ripples in the disc. 

On the other hand, the satellite in a retrograde orbit with $R_i = 6R_d$ starts interacting with a region of the primary with lower density, which leads to a slower orbital decay. After approximately $4\tau$ of evolution, the galaxy's inner region is not significantly affected because the satellite is still not passing close enough to disturb it. 

For the prograde coplanar orbit with $R_p = R_d$ (Orbit 3), a spiral pattern is clearly formed in the gaseous component, though it is not as symmetric as in the previous cases. In the retrograde sense of this orbit, the first passage leaves a trail in the gaseous component that is approximately coincident with the orbit of the satellite as it passes through the disc. This trail eventually rotates with the galaxy. As the orbit of the satellite decays, it continues to produce some damage on the gas disc.

In the simulations with the barred-galaxy (Model B), a prominent $m = 2$ spiral pattern is also produced in both the stellar and gaseous components by the satellite in Orbit 2 in a prograde sense (right panels of Figures \ref{fig:morph-orbit2-pro}). The morphology is similar to that of the simulation with Model A. For the smaller orbit ($R_i = 3 R_d$), noticeable disturbances of the gaseous and stellar components are generated. The more massive satellite produces more prominent effects. In the case of the retrograde orbits, the orbit with $R_i = 3R_d$ is the one that produces the most noticeable morphological effects, as shown in the right panels of Figure \ref{fig:morph-orbit1-ret}. Nevertheless, it is difficult to determine if this is a result of the interaction as these are barely distinguishable from the spiral pattern formed in isolated evolution. As with Model A, no significant effects are observed in the simulation with retrograde Orbit 2 ($R_i = 6R_d$). Finally, in a retrograde coplanar orbit with $R_p = R_d$, the satellite produces noticeable effects on the gas and stellar components.

The above results show that low-mass satellites (${\cal R} \approx 1:1000 - 1:100$) still produce significant morphological changes on the host galaxy. The results are similar in both the non-barred and barred galaxies, but being more prominent with the more massive satellite ($1.2\times 10^{10} \textrm{M}_\odot$).

%%%%%%%%%%%%%%%%%%%%%%%%%%%%%%%%%%%%%%%%%%%%%%%%%%
%%%%%%%%%%%%%%%%%%%%%%%%%%%%%%%%%%%%%%%%%%%%%%%%%%
\subsection{Induced Gas Flows}
\label{subsec:mass-distribution}

The cumulative gas mass function $M_g(<R)$ at the end of each simulation is computed in order to quantify the gas displacement in the host. Due to gradual tilting of the primary by the interaction, the cumulative mass function is calculated in a coordinate system aligned with the disc's principal axes of the moment of inertia. These were determined using only gas and stellar particles with $R < 4R_D$ to exclude any tidal tail structure from this calculation. This allows to obtain $M_g(<R)$ in a plane approximately coincident with the disc's plane. For convenience, $M_g(<R)$ is normalised to the galaxy's total gas mass, and it is expressed as $\mu_g(<R) = M_g(<R)/M_g$.

The following sections, \S \ref{subsubsec:results-flows-modelA} and \ref{subsubsec:results-flows-modelB}, show the plots of $\mu_g(<R)$ as a function of galactocentric radius $R$ for all the simulations at $t \approx 4 \tau$ with the non-barred (Model A) and barred (Model B) galaxies, respectively.
The final gas mass distribution of each simulation is compared with that of the isolated host galaxy at the same dynamical time. Because the bar in Model B drives radial gas flows in isolated evolution, this comparison allows to distinguish the extent of the satellite's effect in producing additional inflows.

%%%%%%%%%%%%%%%%%%%%%%%%%%%%%%%%%%%%%%%%%%%%%%%%%%
\subsubsection{Non-barred Model}
\label{subsubsec:results-flows-modelA}

%%%%%%%%% figs .. Model A 

\begin{figure*}
	\begin{center}
		\includegraphics[width=0.40\linewidth,trim=10mm 5mm 10mm 10mm,clip=true]{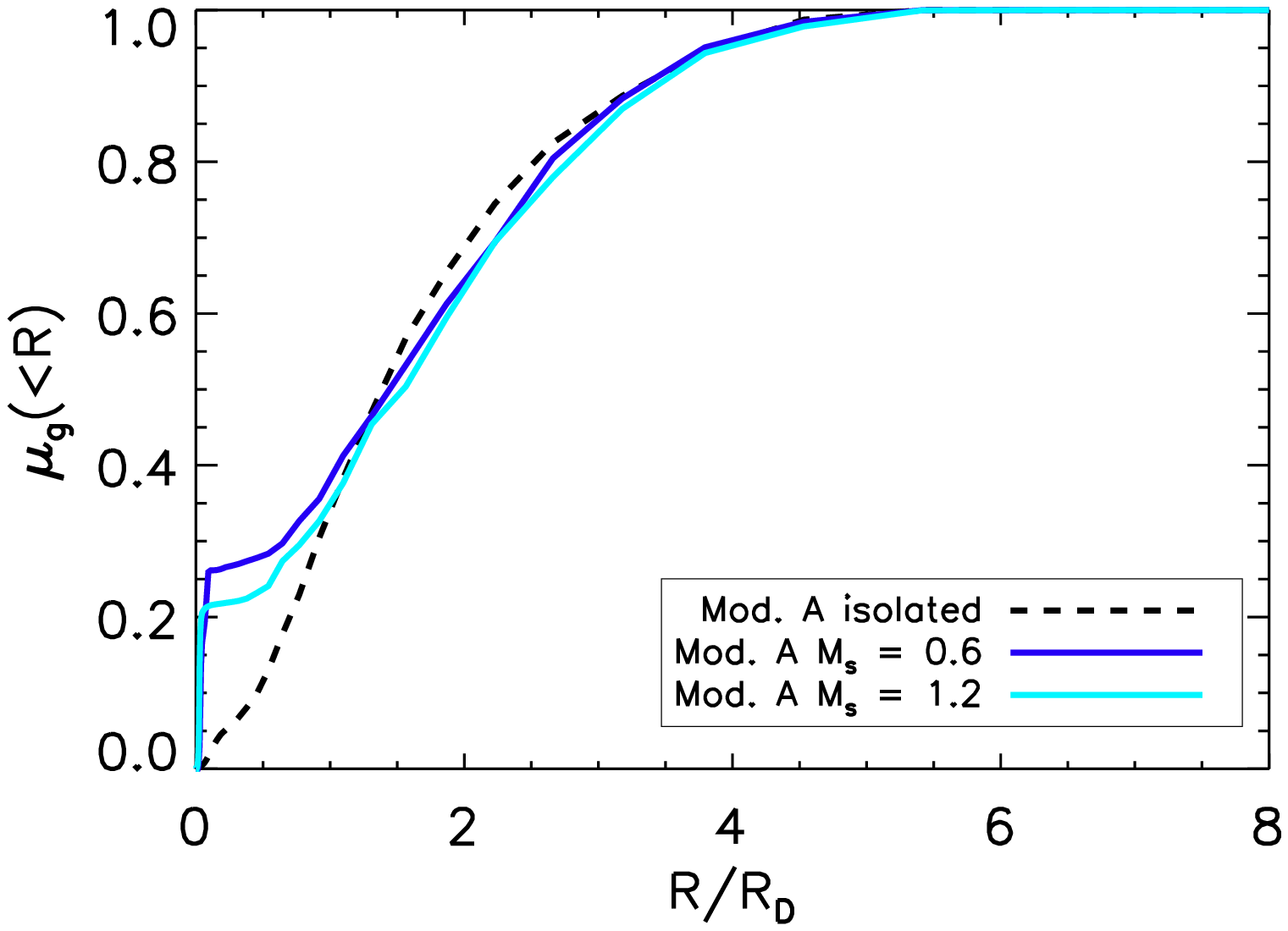}
		\includegraphics[width=0.40\linewidth,trim=10mm 5mm 10mm 10mm,clip=true]{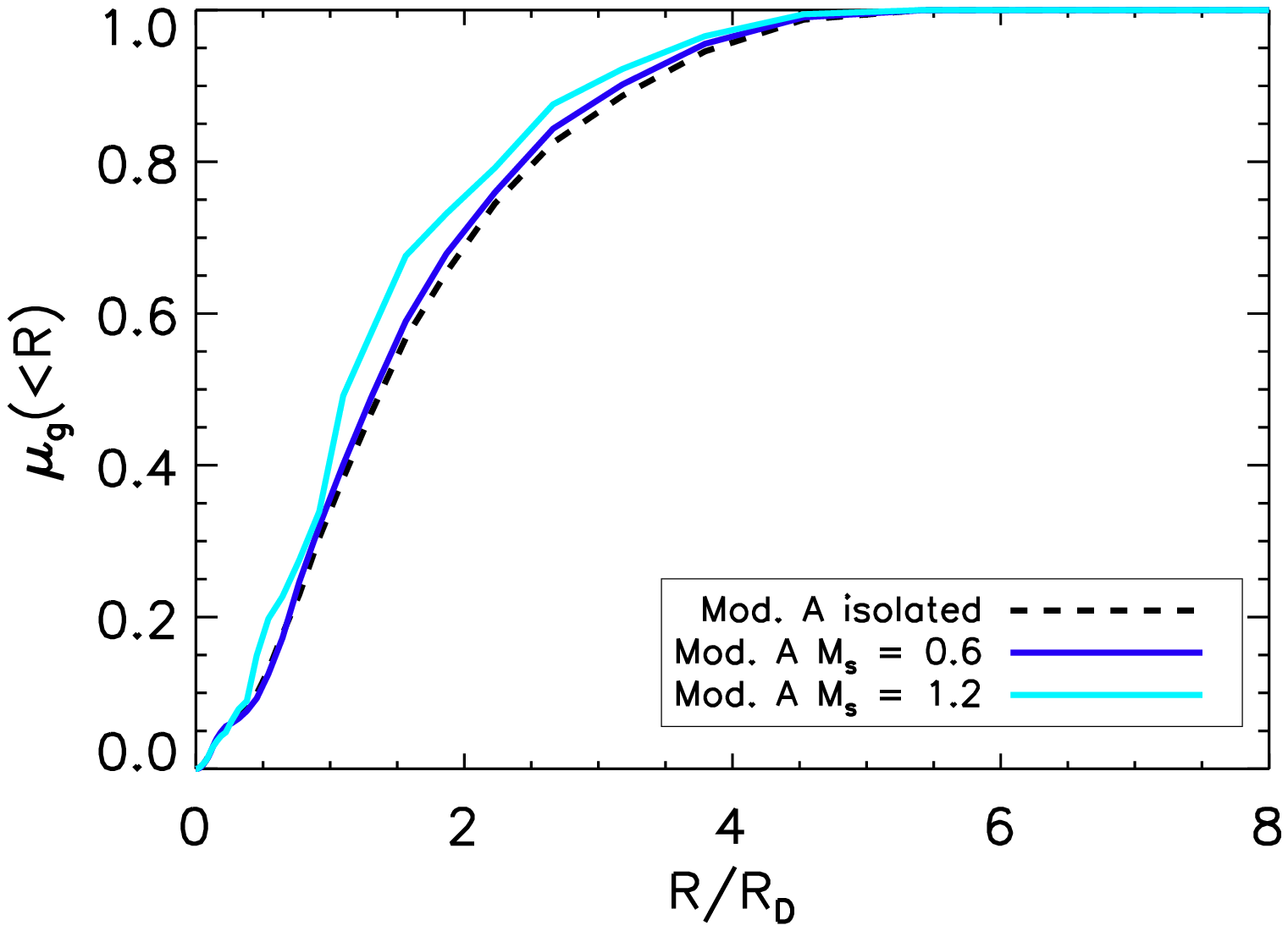}

		\includegraphics[width=0.40\linewidth,trim=10mm 5mm 10mm 10mm,clip=true]{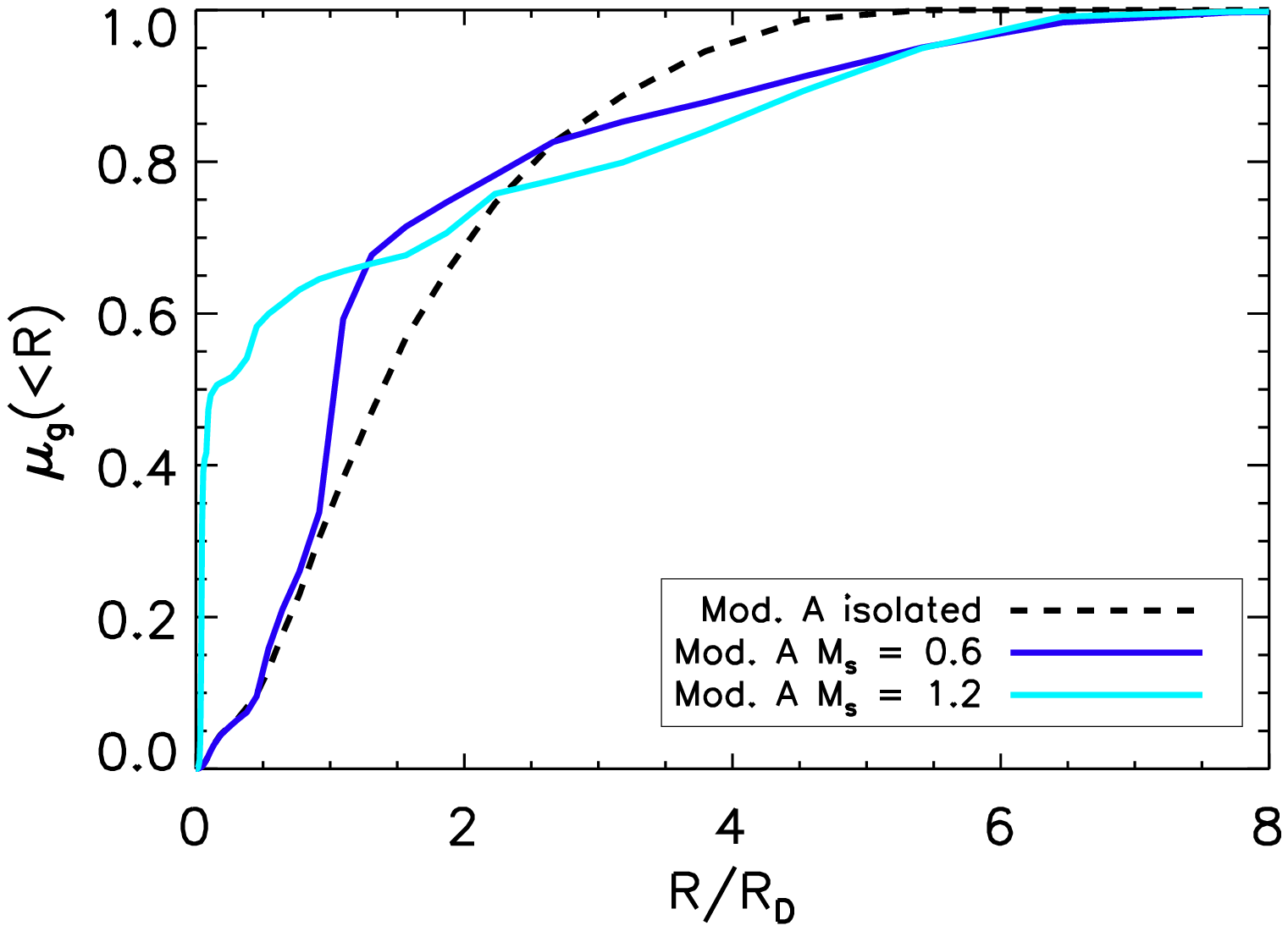}
		\includegraphics[width=0.40\linewidth,trim=10mm 5mm 10mm 10mm,clip=true]{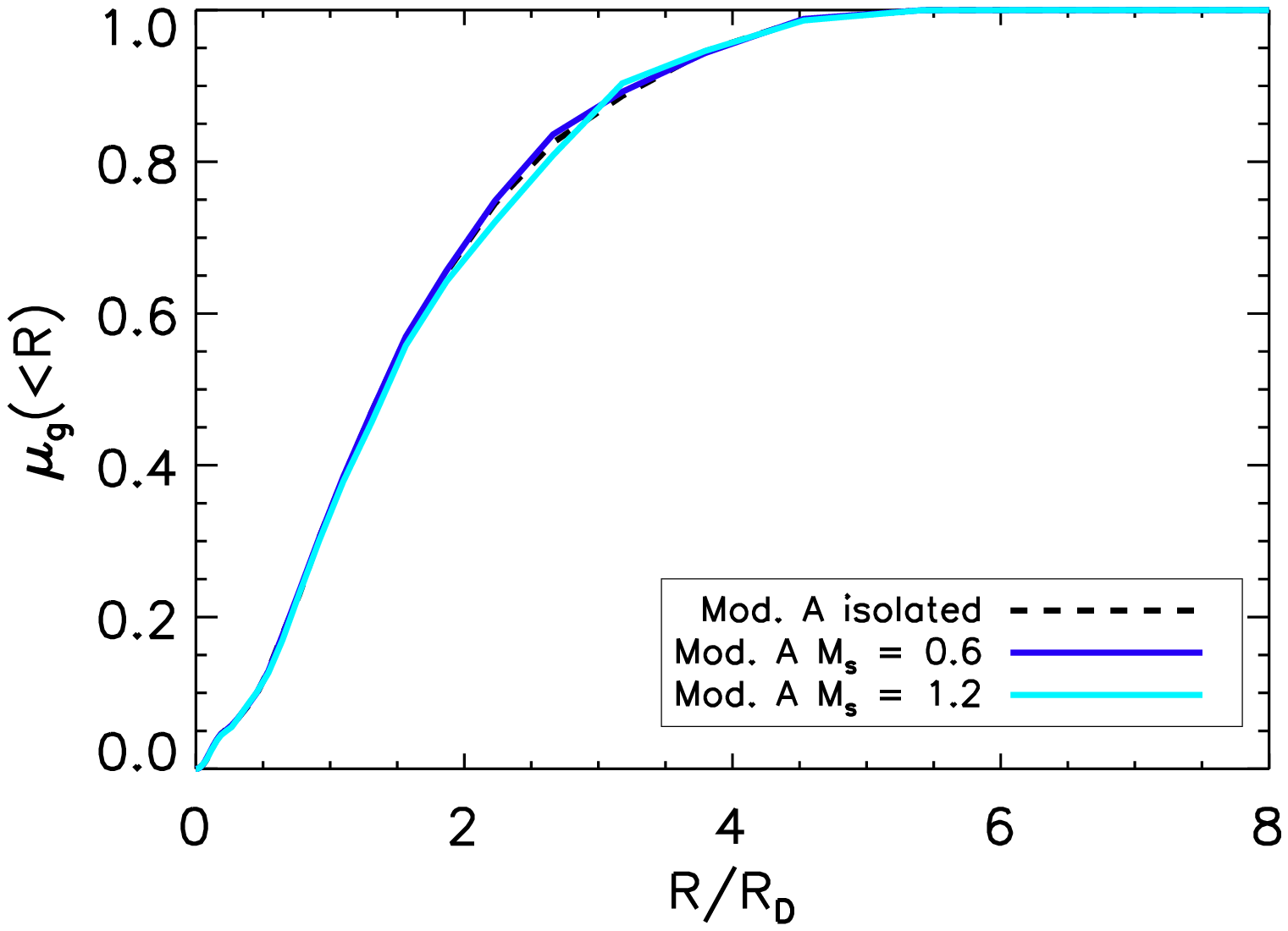}

		\includegraphics[width=0.40\linewidth,trim=10mm 5mm 10mm 10mm,clip=true]{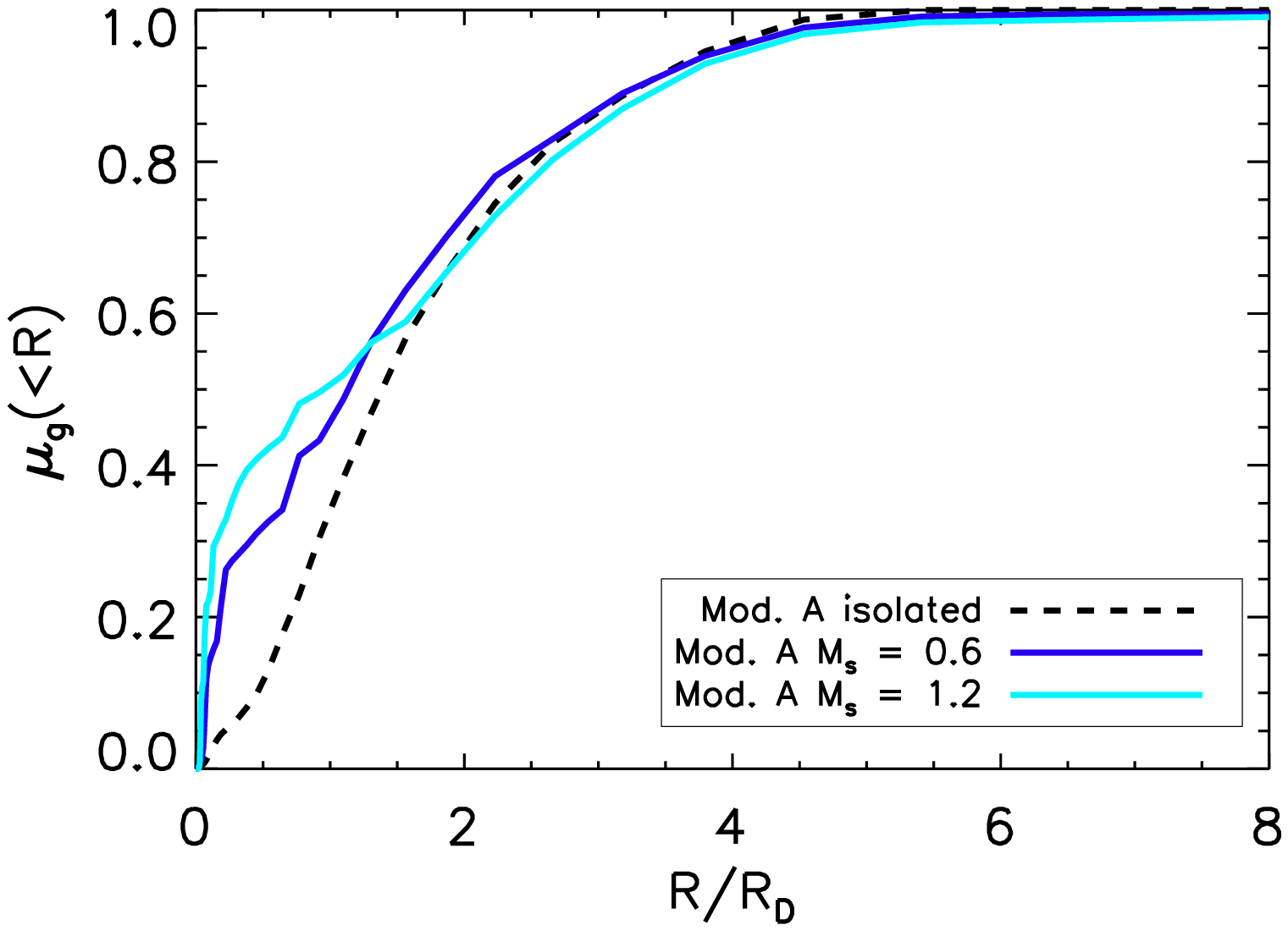}
		\includegraphics[width=0.40\linewidth,trim=10mm 5mm 10mm 10mm,clip=true]{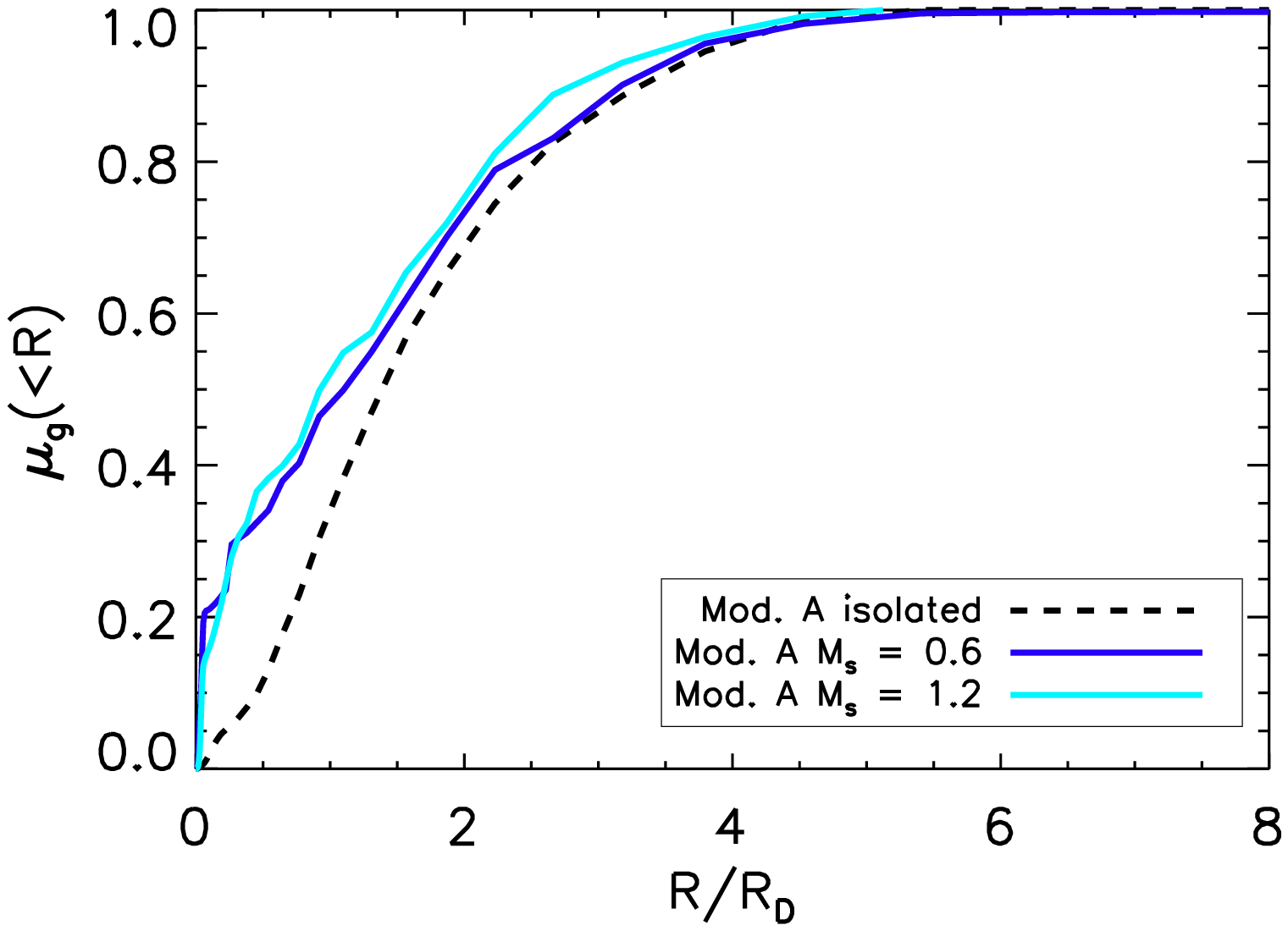}

		\caption{Final integrated mass fraction of the simulations of the non-barred galaxy (Model A) with Orbit 1 ($R_i = 3R_D,\, i = 30^{\circ}$) in the top panel; with Orbit 2 ($R_i = 6R_D,\, i = 30^{\circ}$) in the middle panel; and with Orbit 3 ($R_p = 6R_D,\, i = 0^{\circ}$) in the lower panel. The \emph{left column} shows the results for the prograde orbits, which show an increase in the amount of gas in the central region of the galaxy in the three cases. The \emph{right panel} corresponds to the retrograde orbit, which show no significant difference with respect to the isolated evolution except for the coplanar orbit (bottom right panel). In the latter, it is interesting that gas is displaced by the effect of a retrograde orbit.} 
		\label{fig:mass-distributions-model-A}
	\end{center}
\end{figure*}

The final gas mass distribution $\mu_g(<R)$ for the prograde encounter of the satellite with the non-barred galaxy (Model A) in Orbit 1 ($R_i = 3R_D,\, i = 30^{\circ}$) is shown in the top left panel of Figure~\ref{fig:mass-distributions-model-A}. Although $\mu_g(<R_D)$ increases to $\approx 0.34$, compared to $\mu_g(<R_D) = 0.30$ for the isolated case, Figure~\ref{fig:mass-distributions-model-A} does show a significant increase in $\mu_g$ in the inner region of the galaxy. This indicates that radial gas flows have transferred mass to $R < R_D$.

At $R > 1.5R_D$, the final $\mu_g$ is slightly lower with respect to the isolated case, showing that gas has also moved to higher radii. This may be a result of the energy and momentum transferred by the satellite during its first passages through the disc. The gas distribution produced by the more massive satellite is not significantly different to that of the less massive satellite. At small radii, the final cumulative mass distribution produced by the $M_s = 1.2$ (Satellite 2) is $\approx 8$ \% lower than that of the satellite with $M_s = 0.6$ (Satellite 1).

In the retrograde case of this orbit, the top right panel of Figure \ref{fig:mass-distributions-model-A} shows that the more massive satellite produces a slightly displaced mass distribution compared to that of isolated evolution. This shows that a satellite may induce flows even in retrograde encounters, though this effect may be enhanced by the particular orbit chosen in this work. However, the gas is not reaching the central regions as it happens in the prograde case. The final $\mu_g(<R)$ profile has a shape similar to that of the isolated profile but with a steeper slope. The effect of the less massive satellite is almost indiscernible from the final $\mu_g$ in isolated evolution.

For Orbit 2 ($R_i = 6R_D,\, i = 30^{\circ}$), the results of a prograde encounter with Model A are shown in the middle left panel of Figure \ref{fig:mass-distributions-model-A}. It shows an important difference between the final mass distribution of Satellite 1 and that of Satellite 2. The more massive satellite (Satellite 2) drives a significant amount of gas to the galaxy's central region: at $R = R_D$, $M(<R_D) \approx 0.63 M_g$. This is an increase of about a factor of 2 with respect to isolated evolution. For the less massive satellite, the amount of displaced gas decreases. At $R = R_D$, $M(<R_D) \approx 0.36M_d$, which is $\approx 10 \%$ more than the value in isolated evolution. At $R = 2R_D$, the final distribution increases by $\approx 14 \%$ with respect to the isolated case. 

At $R > 2R_D$, both distributions fell with respect to that of isolated evolution, implying that gas is also being driven to larger radii by the interaction with the satellite. This may be explained by the fact that the satellite first disturbs the outer regions of the galaxy, which have a lower density and are less bound. Particles in these regions can be driven to higher orbits as the satellite passes through.

The middle right panel of Figure \ref{fig:mass-distributions-model-A} shows the final mass distributions for the retrograde encounters in Orbit 2. There is no significant difference between the final $\mu_g$ produced by the satellite and that of isolated evolution as well as between the distributions produced by each satellite. Because the orbital decay takes longer for this orbit, the satellite remains for a longer period disturbing the outer parts of the galaxy, which explains the lack of important changes in the primary's gas distribution.

The final $\mu_g(<R)$ for the simulations with Orbit 3 ($R_p = R_D,\, i = 30^{\circ}$) are shown in the bottom panels of Figure \ref{fig:mass-distributions-model-A}. The left panel shows the results of the prograde orbit and the right one shows those of the retrograde orbit. Both prograde and retrograde encounters produce final mass distributions with a higher mass fraction in the inner regions of the disc. For the prograde case (bottom left panel of Figure \ref{fig:mass-distributions-model-A}), results show that Satellite 2 drives a slightly higher amount of gas to the central region than Satellite 1. At $R = R_D$, $\mu_g(<R)$ in the simulation with Satellite 1 is $\approx 42 \%$ higher than that of the isolated case, and in the simulation with Satellite 2, it is $\approx 58 \%$ higher than in isolated evolution. For the retrograde case, the final distributions are similar for both satellites. Satellite 1 produces an increase in the enclosed mass at $R = R_D$ of $\approx 52 \%$, and Satellite 2 produces an increase of about $\approx 60\%$. Although Orbit 3 is expected to alter the gas distribution in both prograde and retrograde orientations, it is interesting to note that for Model A the change in $\mu_g$ still shows some dependence on the satellite's mass. We present some resolution tests using Model A in \S \ref{sec:A-Res-Tests}, which show that $\mu_g(<R)$ is not strongly sensitive to the resolution.

%%%%%%%%%%%%%%%%%%%%%%%%%%%%%%%%%%%%%%%%%%%%%%%%%%
%%%%%%%%%%%%%%%%%%%%%%%%%%%%%%%%%%%%%%%%%%%%%%%%%%
\subsubsection{Barred Model}
\label{subsubsec:results-flows-modelB}

%%%%%%%%% figs .. Model B

\begin{figure*}
	\begin{center}
		\includegraphics[width=0.40\linewidth,trim=10mm 5mm 10mm 10mm,clip=true]{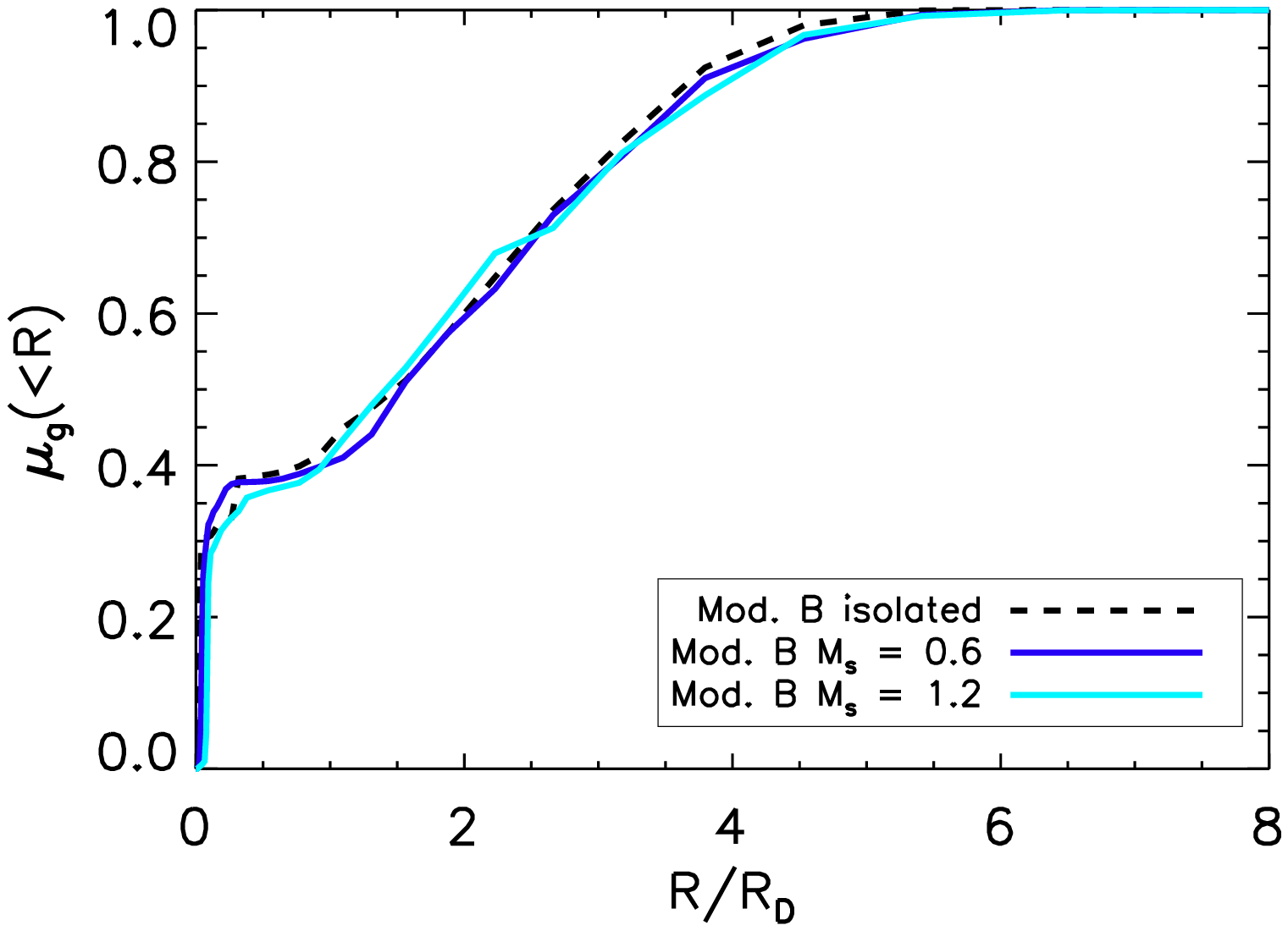}
		\includegraphics[width=0.40\linewidth,trim=10mm 5mm 10mm 10mm,clip=true]{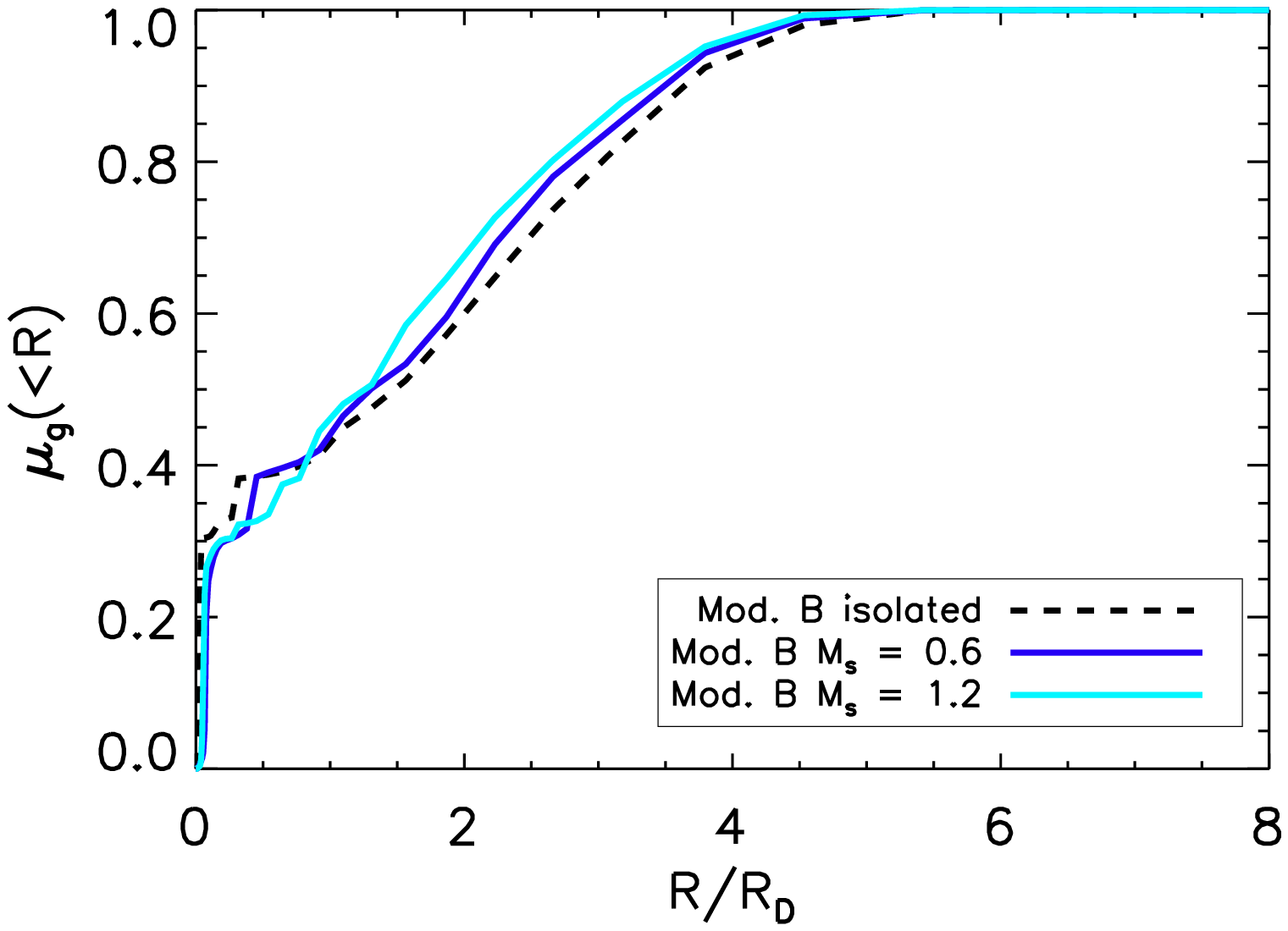}

		\includegraphics[width=0.40\linewidth,trim=10mm 5mm 10mm 10mm,clip=true]{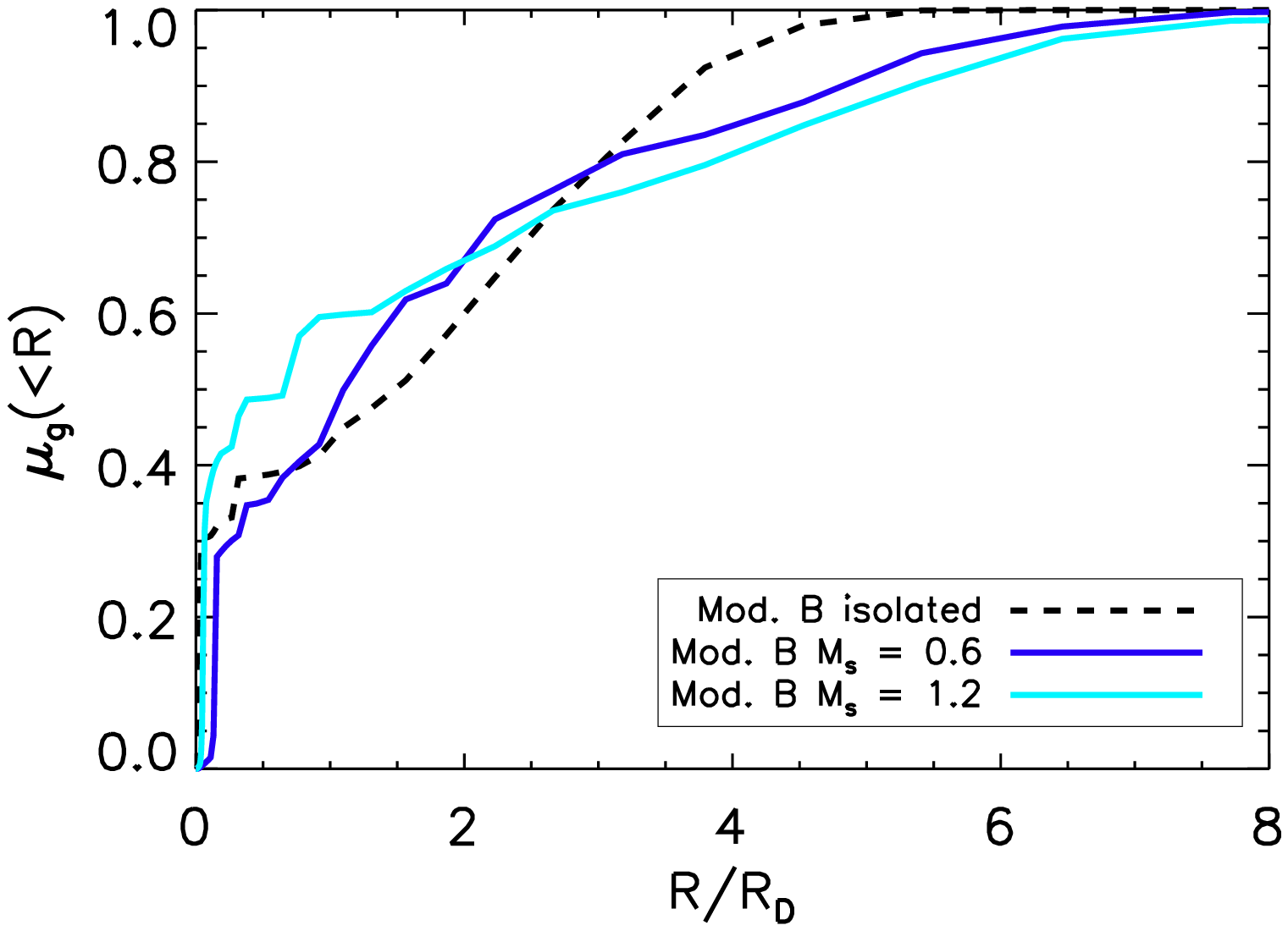}
		\includegraphics[width=0.4\linewidth,trim=10mm 5mm 10mm 10mm,clip=true]{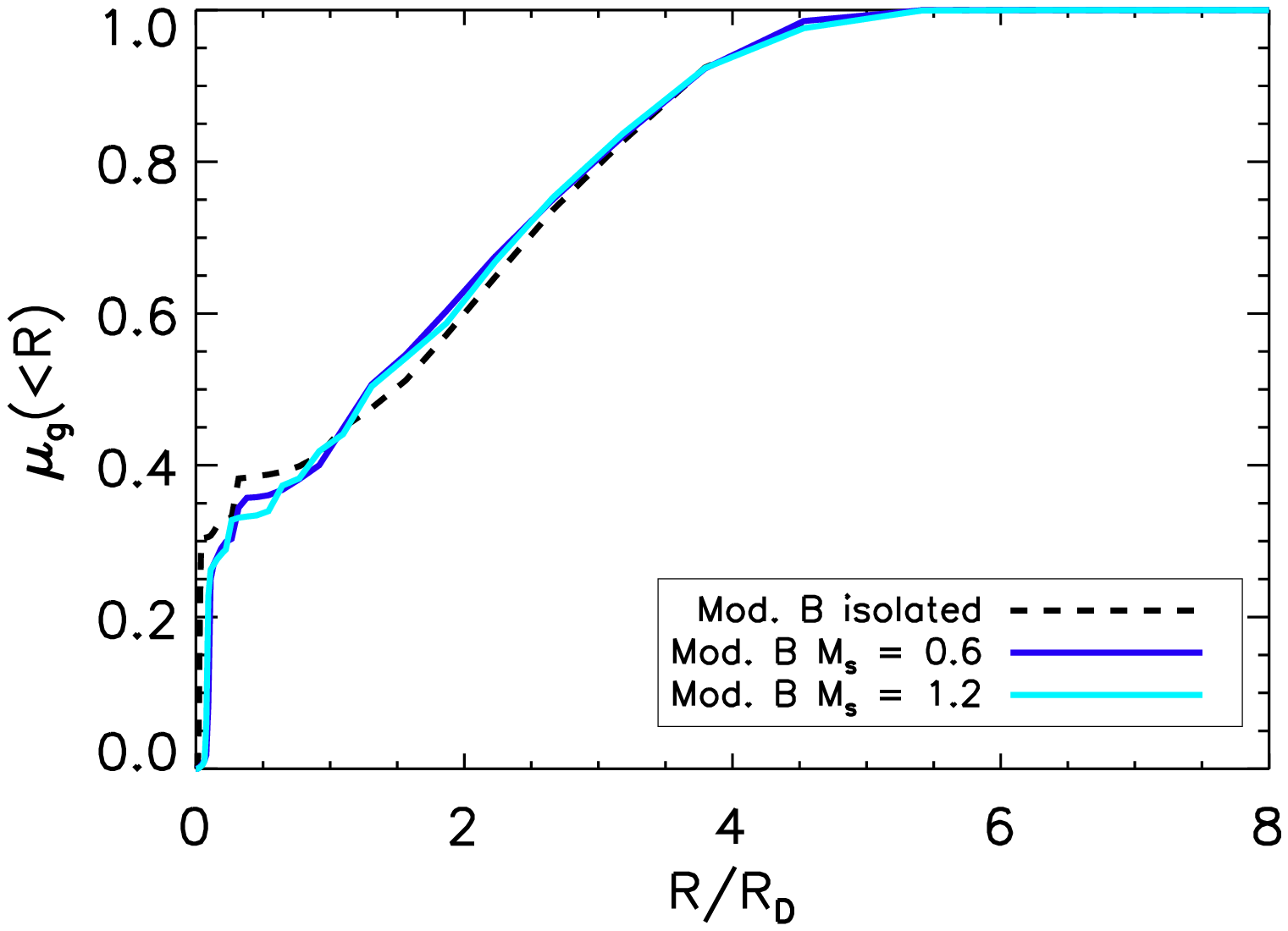}

		\includegraphics[width=0.40\linewidth,trim=10mm 5mm 10mm 10mm,clip=true]{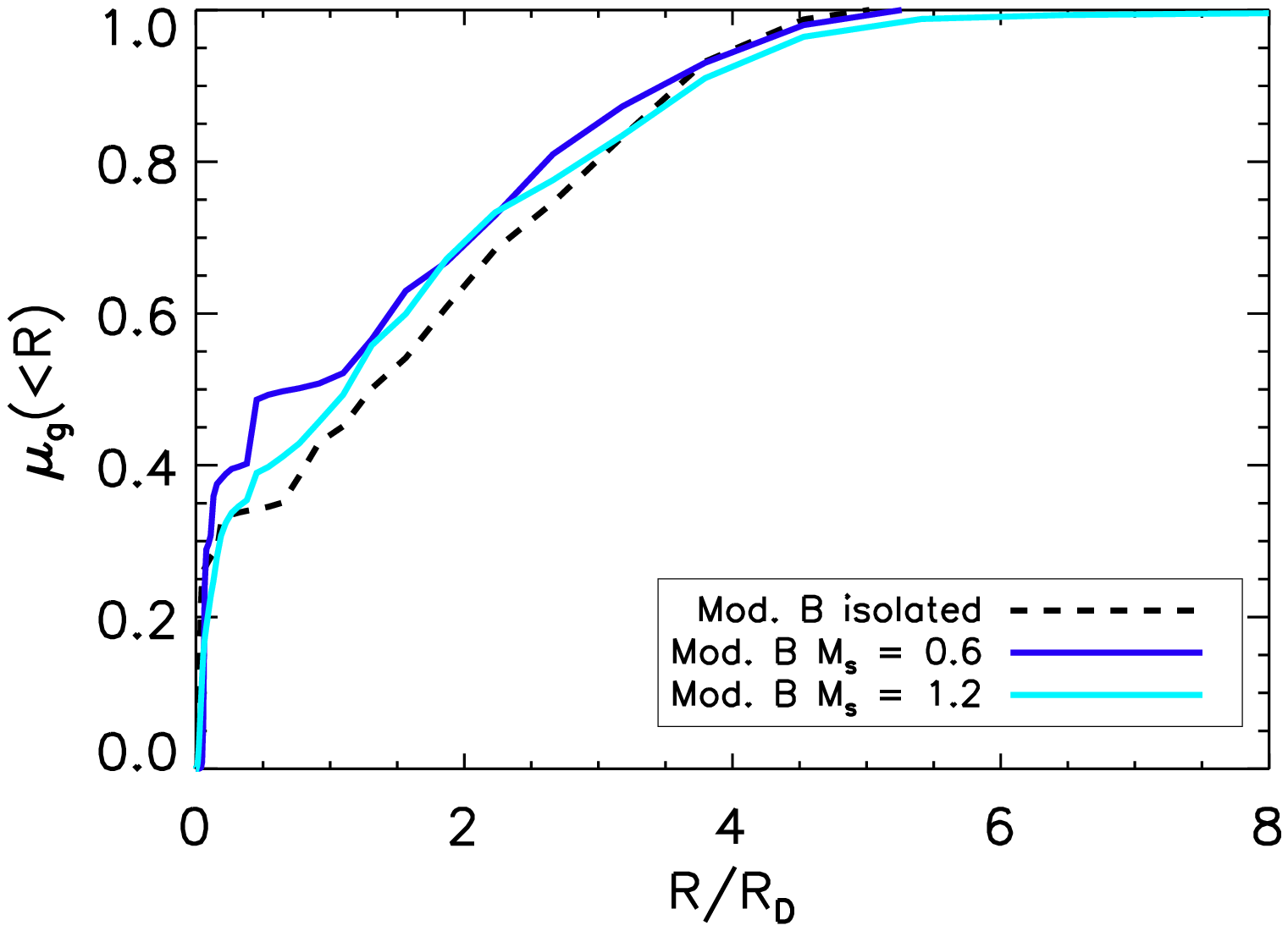}
		\includegraphics[width=0.40\linewidth,trim=10mm 5mm 10mm 10mm,clip=true]{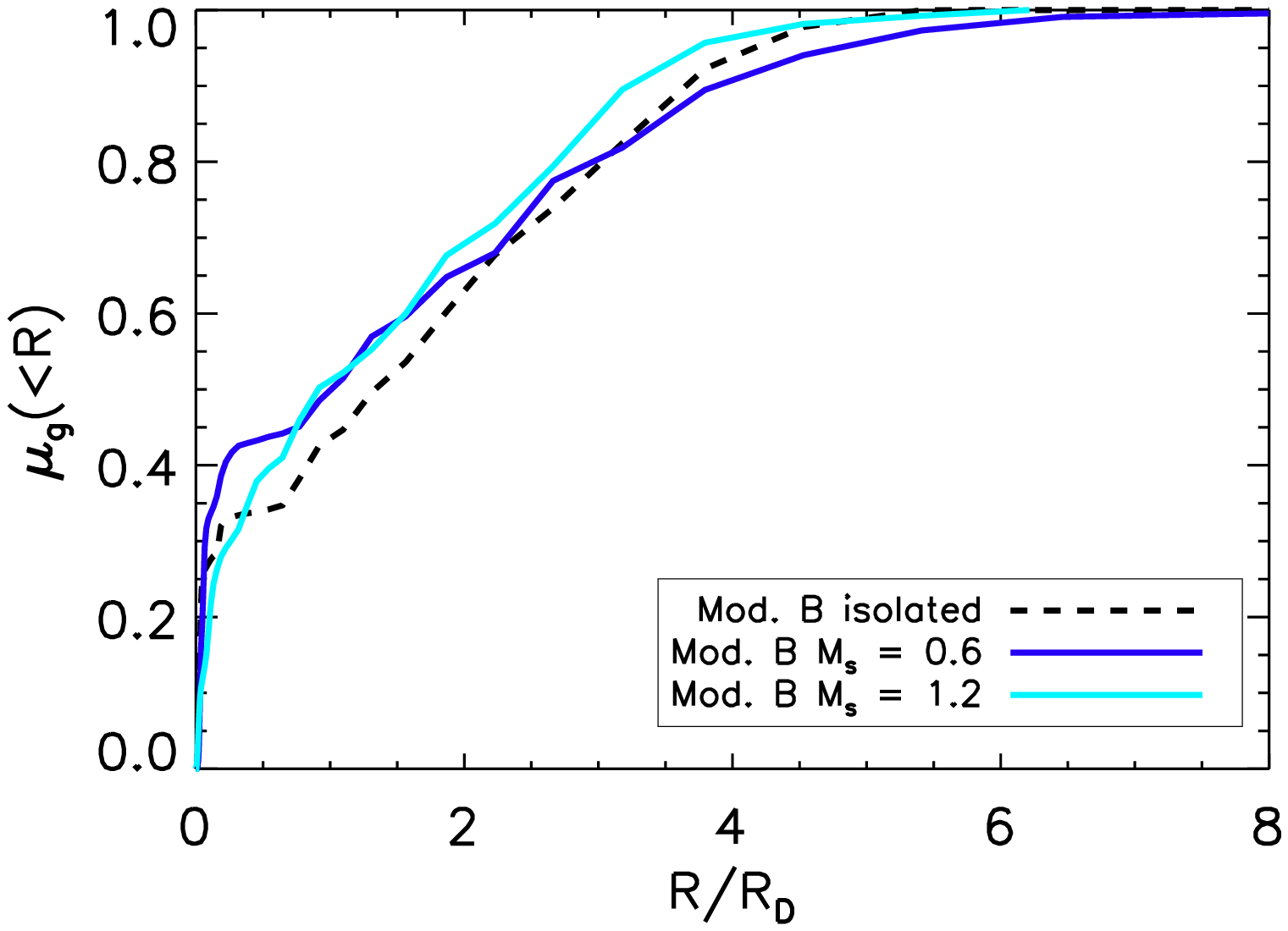}
		\caption{Final integrated mass fraction of the simulations of the barred galaxy (Model B) with Orbit 1 ($R_i = 3R_D,\, i = 30^{\circ}$) in the top panels; with Orbit 2  ($R_i = 6R_D,\, i = 30^{\circ}$) in the middle panels; and with Orbit 3 ($R_p = 6R_D,\, i = 0^{\circ}$). The \emph{left column} shows the results of the simulations with the prograde orbits and the \emph{right column} presents the ones with the retrograde orbits. The changes in the mass distribution tend to be less significant compared to the isolated model as well as to Model A.}
		\label{fig:mass-distributions-model-B}
	\end{center}
\end{figure*}

In the simulations with Model B, a prograde encounter in Orbit 1 produces the cumulative mass functions shown in the top left panel of Figure \ref{fig:mass-distributions-model-B}. The difference in $\mu_g$ between the interacting pair and the isolated model is not significant. At $R = R_D$, for both satellites $\mu_g(<R)$ is approximately $4 \%$ lower than the gas fraction in isolated evolution. For the retrograde case, shown in the top right panel of Figure \ref{fig:mass-distributions-model-B}, the slope of the final cumulative mass function is higher, which is indicative of radial inflows. This tendency is more evident for the more massive satellite (Satellite 2). At $R = R_D$, $\mu_g$ is $\approx 2 \%$ higher for Satellite 1, and about $8 \%$ for Satellite 2 with respect to isolated evolution. At $R = 2R_D$, this difference is $\approx 13 \%$ for the simulation with the second satellite. These results show a tendency similar to those of Model A.

For the simulations with Orbit 2, the middle left panel of Figure \ref{fig:mass-distributions-model-B} shows the final gas mass distributions for the prograde encounter. In this orbit, the more massive satellite displaces a much higher amount of gas to the central region than Satellite 1. At $R = R_D$, $\mu_g$ for the simulation with Satellite 1 is about $3 \%$ higher than that of the isolated case; at $R = 2R_D$, $\mu_g$ is $10 \%$ higher. For the more massive satellite, $M(<R) = 0.60M_g$ at $R = R_d$,  which is $\approx 44 \%$ higher than the final fraction in isolated evolution. At $R = 2R_D$, the integrated mass is similar to that of the less massive satellite simulation. At $R > 2R_D$, the lower value of the mass fraction with respect to isolated evolution may be attributed to the fact that gas has moved to larger orbits. The middle right panel of Figure \ref{fig:mass-distributions-model-B} shows that in a retrograde encounter, the satellite is not introducing any effect distinguishable from isolated evolution. For both satellites, the final $\mu_g(R<R_D)$ is only about $1 \%$ higher than in isolated evolution. The results are similar to those of the simulations of Model A.

In the case of Orbit 3, the final cumulative gas mass fraction is plotted in the bottom panels of Figure \ref{fig:mass-distributions-model-B}. Both prograde and retrograde orbits produce final mass distributions showing that gas moves to the galaxy's inner parts. However, the effect seems not to be strongly dependent on the satellite's mass. For the prograde orbit (left panel), $\mu_g$ at $R = R_D$ is $\approx 21 \%$ higher for the simulation with Satellite 1, and $\approx 8 \%$ higher than isolated evolution for Satellite 2. The difference between the two plots is practically negligible at higher radii. At $R = 2R_D$, the final fraction for both simulations is $\approx 16 \%$ higher than that of isolated evolution. For the retrograde orbit (bottom right panel of Figure \ref{fig:mass-distributions-model-B}), the integrated gas mass at $R = R_D$ increased $\approx 10 \%$ with respect to that of isolated evolution.

The difference between the prograde and retrograde encounters is clear for the inclined orbits (Orbits 1 and 2). The prograde orbits drive a higher amount of gas than the retrograde ones. The strongest effect is produced by the prograde cases of Orbit 2, which shows a dependence on the mass of the satellite. In the case of Orbit 3, both the prograde and retrograde encounters drive gas to within $R < R_D$. The difference between a prograde and retrograde orbit is not significant, but it depends slightly with mass in Model A.

%%%%%%%%%%%%%%%%%%%%%%%%%%%%%%%%%%%%%%%%%%%%%%%%%%
%%%%%%%%%%%%%%%%%%%%%%%%%%%%%%%%%%%%%%%%%%%%%%%%%%
\subsection{Gas Density Distributions}

In this section we briefly describe the effect of the interaction in the gas density distribution since changes in volume density can be indicative of variations star formation activity. We focus on the number density ($n$) distribution.

\begin{figure*}
	\begin{center}
		\includegraphics[width=0.38\linewidth,height=0.285\linewidth,trim=0mm 5mm 0mm 5mm,clip=true]{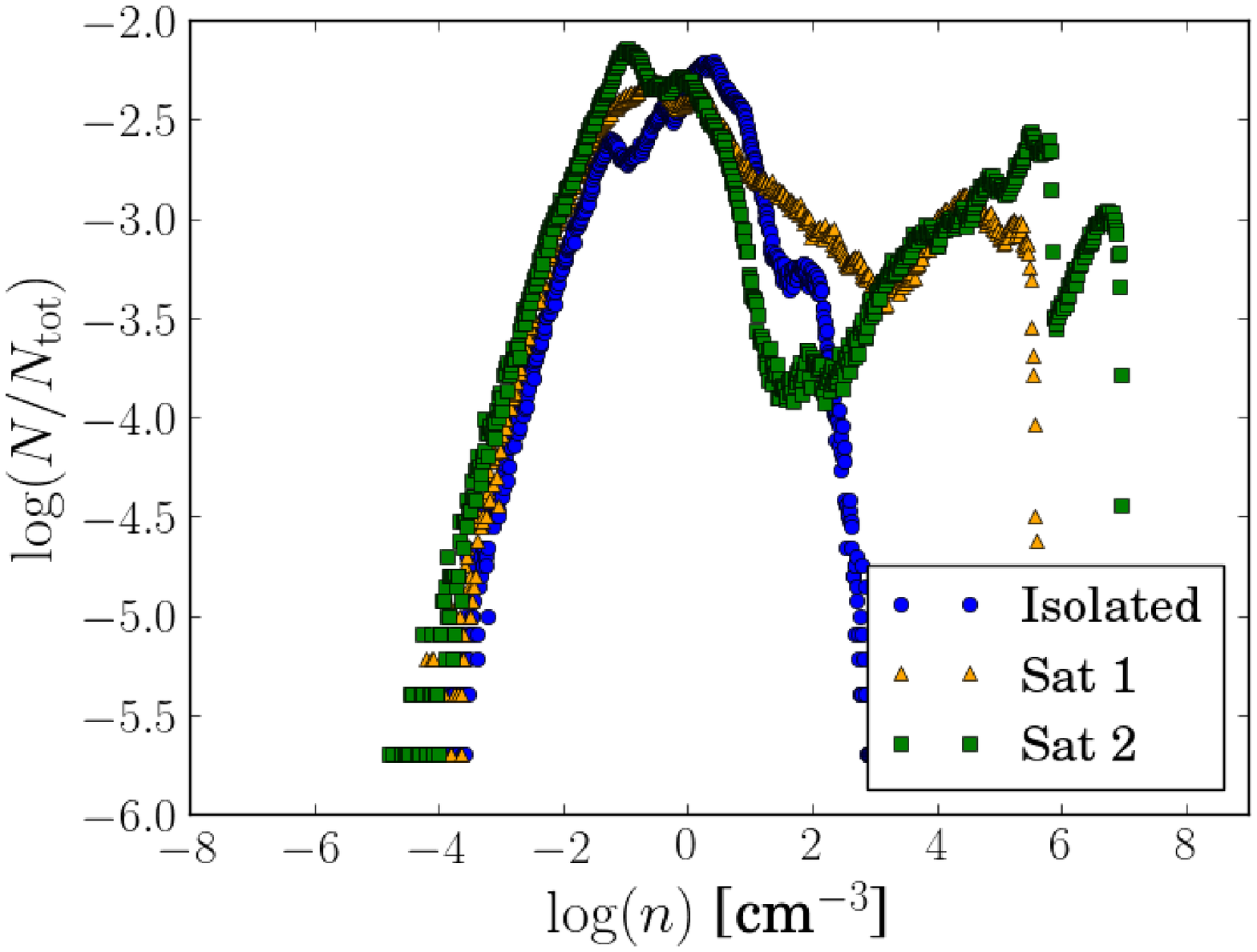}
		\includegraphics[width=0.38\linewidth,height=0.285\linewidth,trim=0mm 5mm 0mm 5mm,clip=true]{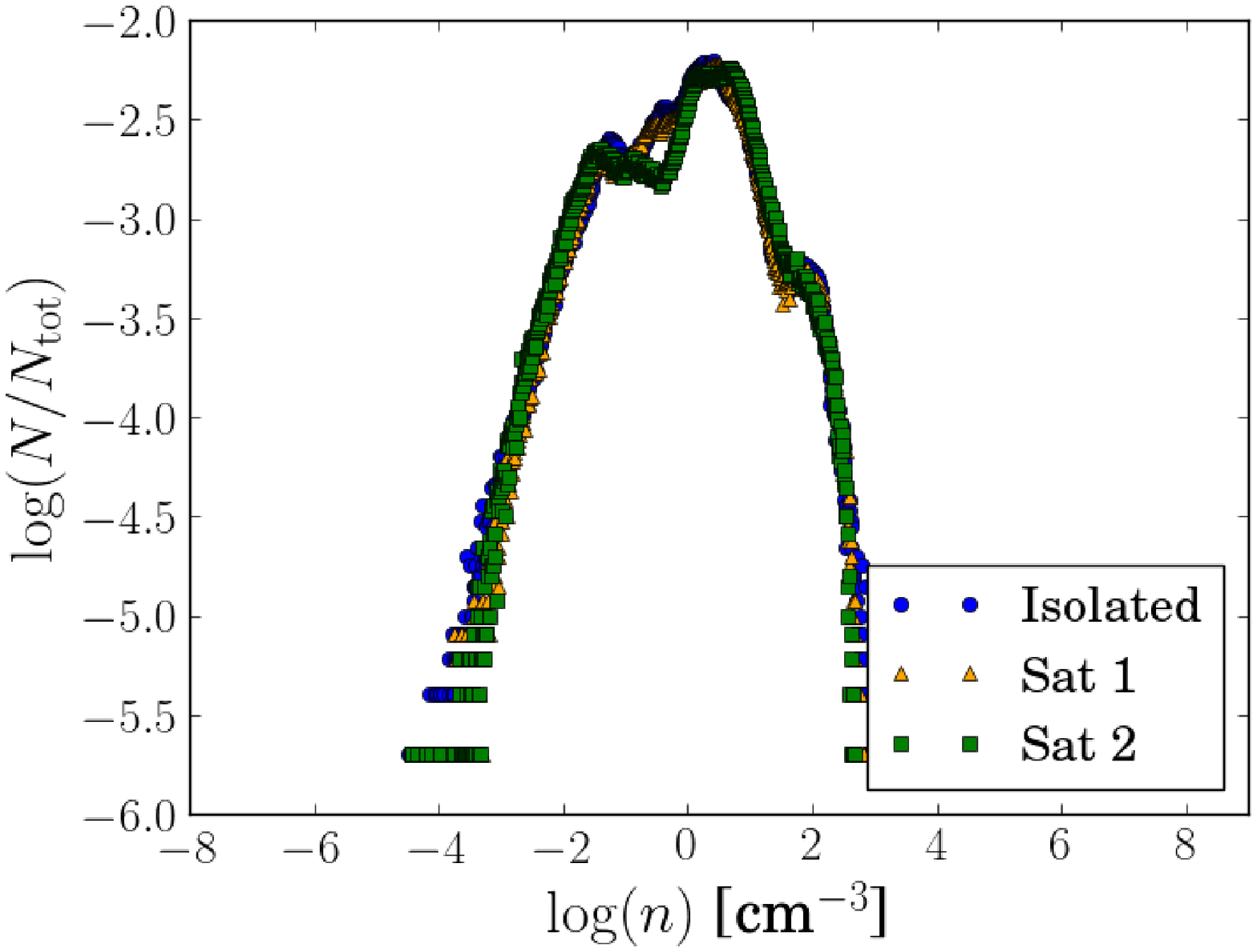}

		\includegraphics[width=0.38\linewidth,height=0.285\linewidth,trim=0mm 5mm 0mm 5mm,clip=true]{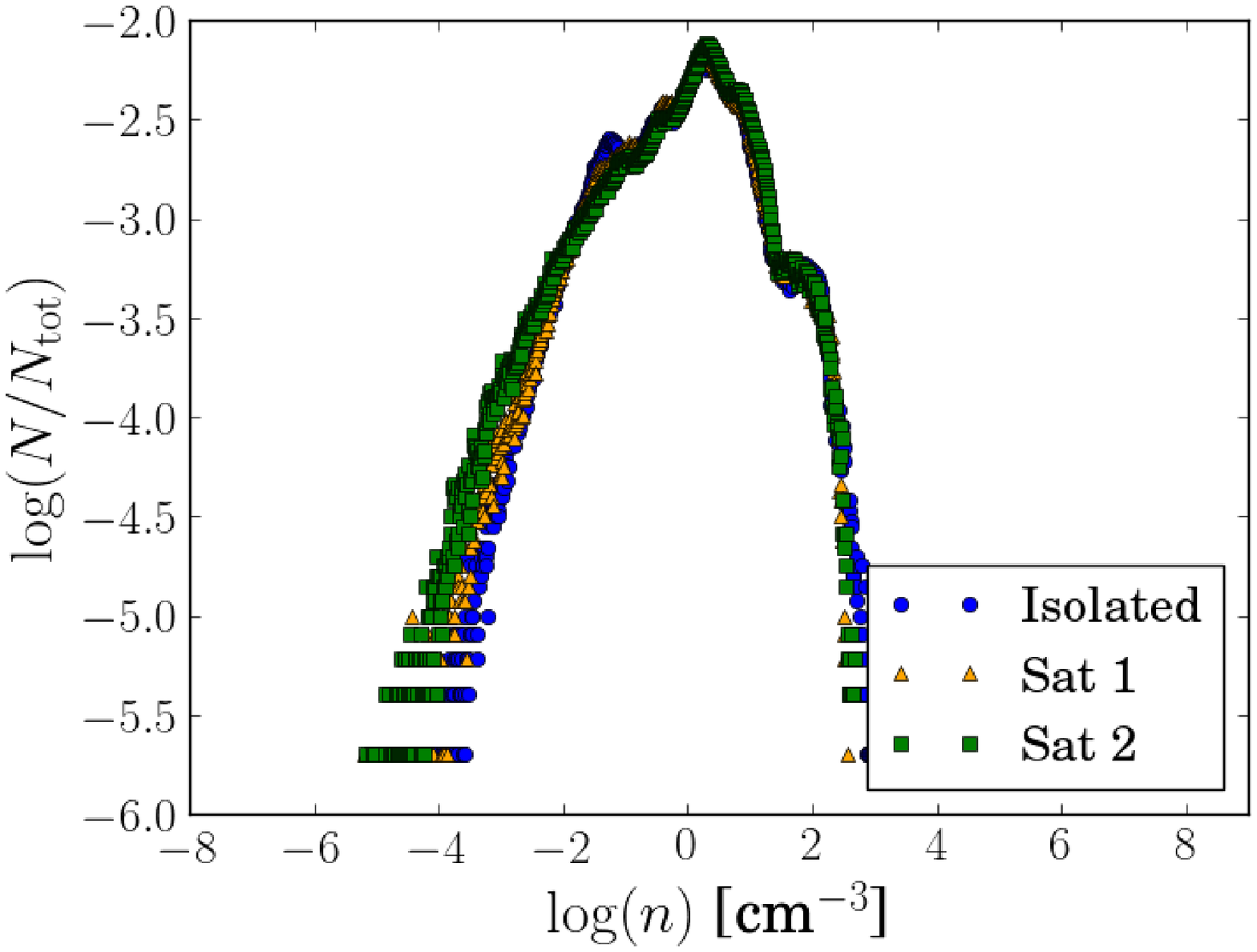}
		\includegraphics[width=0.40\linewidth,height=0.285\linewidth,trim=0mm 5mm 0mm 5mm,clip=true]{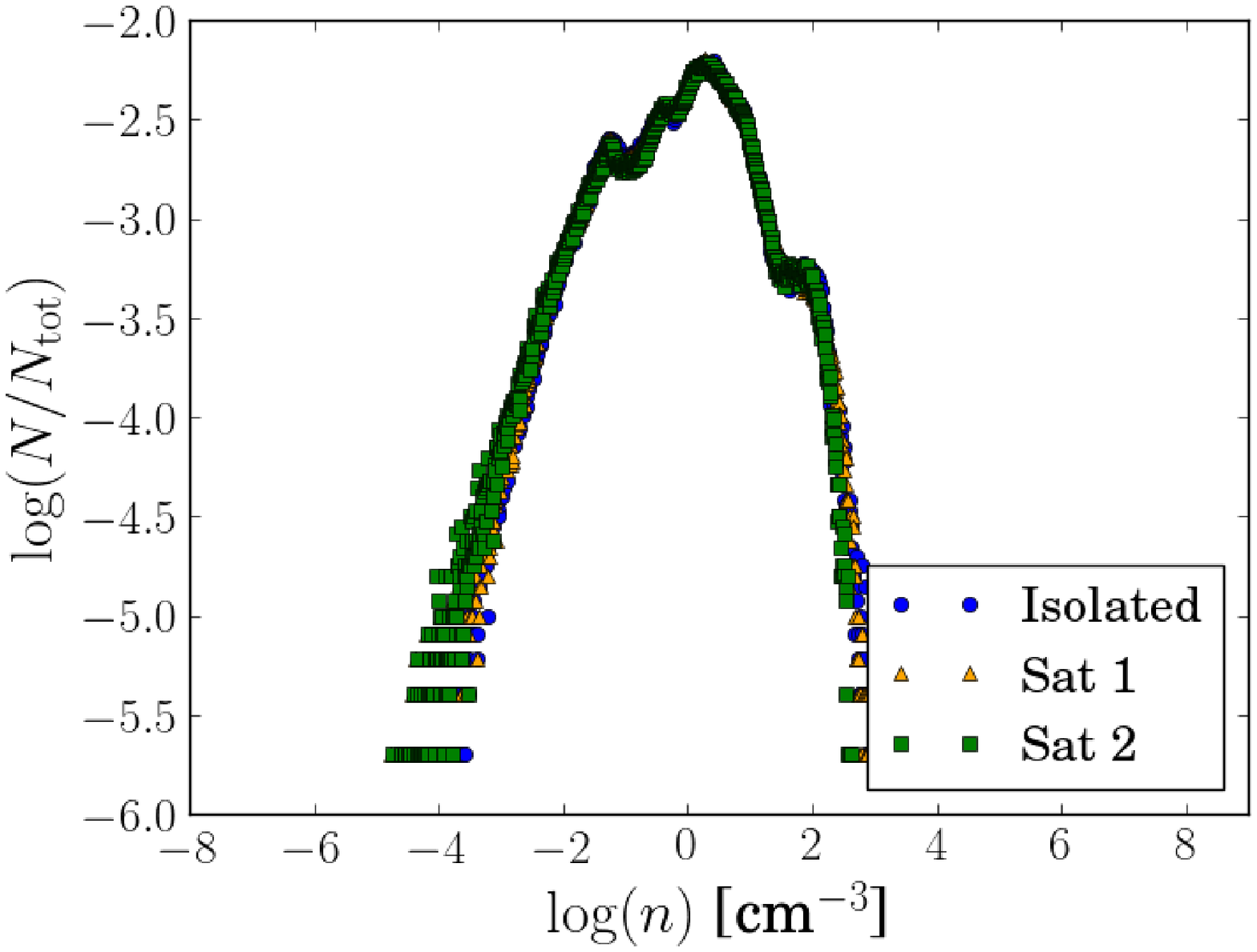}

		\includegraphics[width=0.38\linewidth,height=0.285\linewidth,trim=0mm 5mm 0mm 5mm,clip=true]{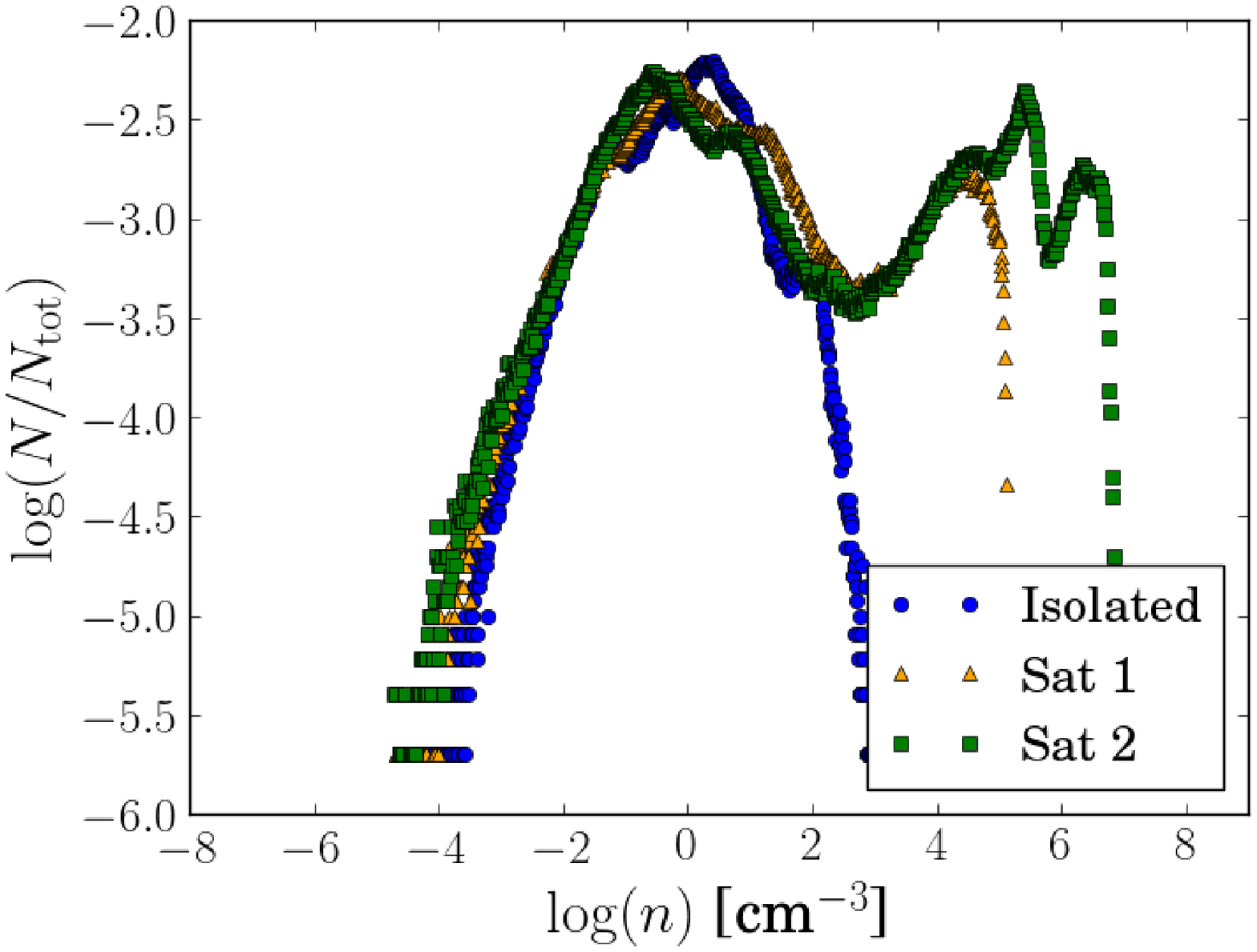}
		\includegraphics[width=0.38\linewidth,height=0.285\linewidth,trim=0mm 5mm 0mm 5mm,clip=true]{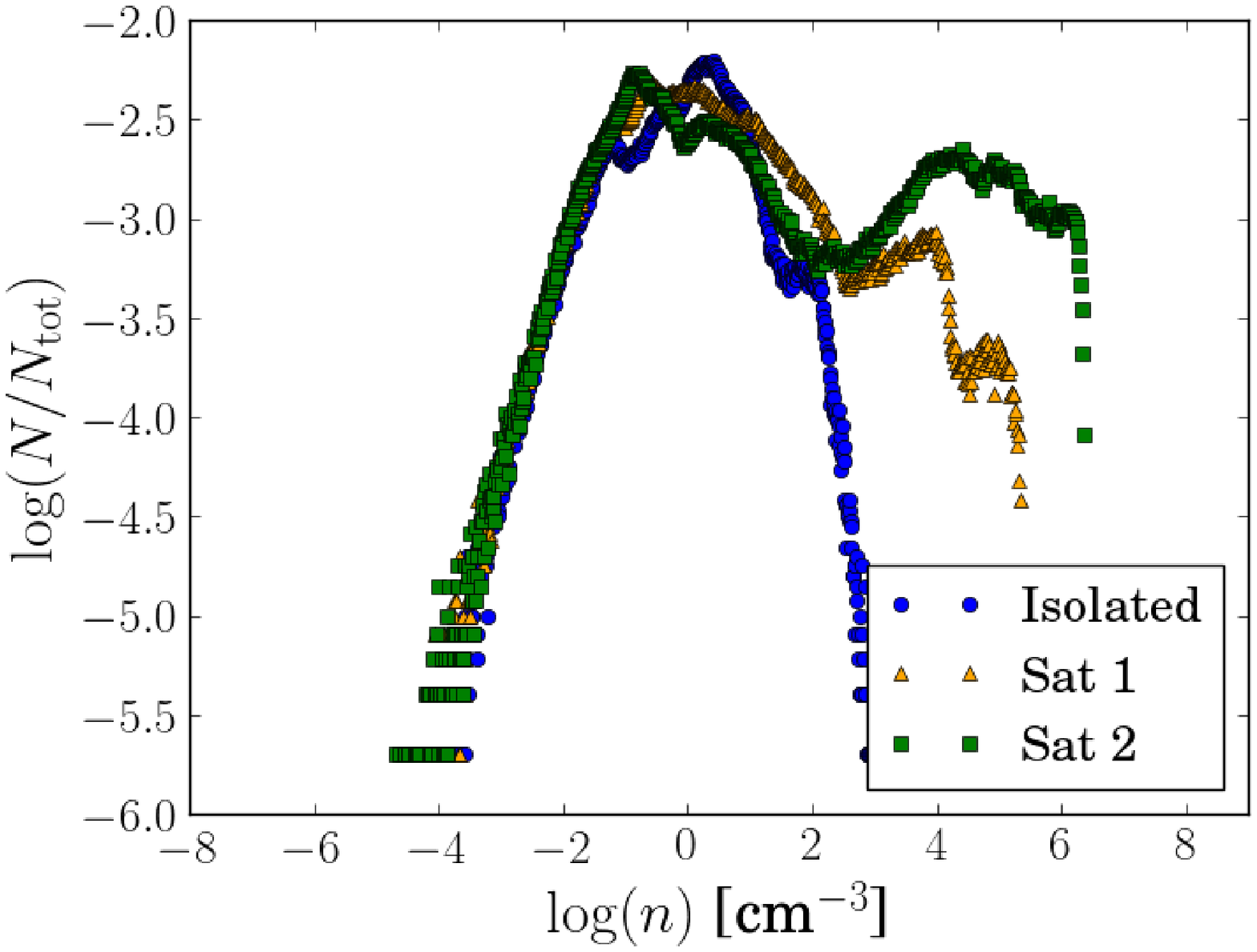}
		\caption{Number density distributions for the simulations of Model A with Orbit 1 ($R_i = 3R_D,\, i = 30^{\circ}$) in the top panels; with Orbit 2  ($R_i = 6R_D,\, i = 30^{\circ}$) in the middle panels; and with Orbit 3 ($R_p = 6R_D,\, i = 0^{\circ}$). The \emph{left column} shows the results of the simulations with the prograde orbits and the \emph{right column} presents the ones with the retrograde orbits. These are calculated at $t \approx 282$ Myr since the beginning of the simulation.}
		\label{fig:dens-dist-mod-A}
	\end{center}
\end{figure*}

In the simulation of Model A and prograde Orbit 1, shown in the top left panel of Figure \ref{fig:dens-dist-mod-A}, the density distribution shows a noticeable increase of particles in higher density bins at $t \approx 282$ Myr from the beginning of the simulation (the first disc passage occurs at $t \approx 100$ Myr). This change is larger for Satellite 2. However, at a later time ($t \approx 470$ Myr), both distributions reach approximately the same maximum densities, with the high-density tail of the distribution for Satellite 2 reaching slightly larger values than the one for the Satellite 1 simulation. For the retrograde version of this simulation, the density distributions are not very different to that of the isolated model at $t \approx 282$ Myr (see the top right panel of Figure \ref{fig:dens-dist-mod-A}). However, at $t \approx 470$ Myr, the distribution for the simulation with Satellite 2 reaches slightly higher densities than the one in isolated evolution.

\begin{figure}
	\begin{center}
		\includegraphics[width=0.93\linewidth,height=0.6975\linewidth]{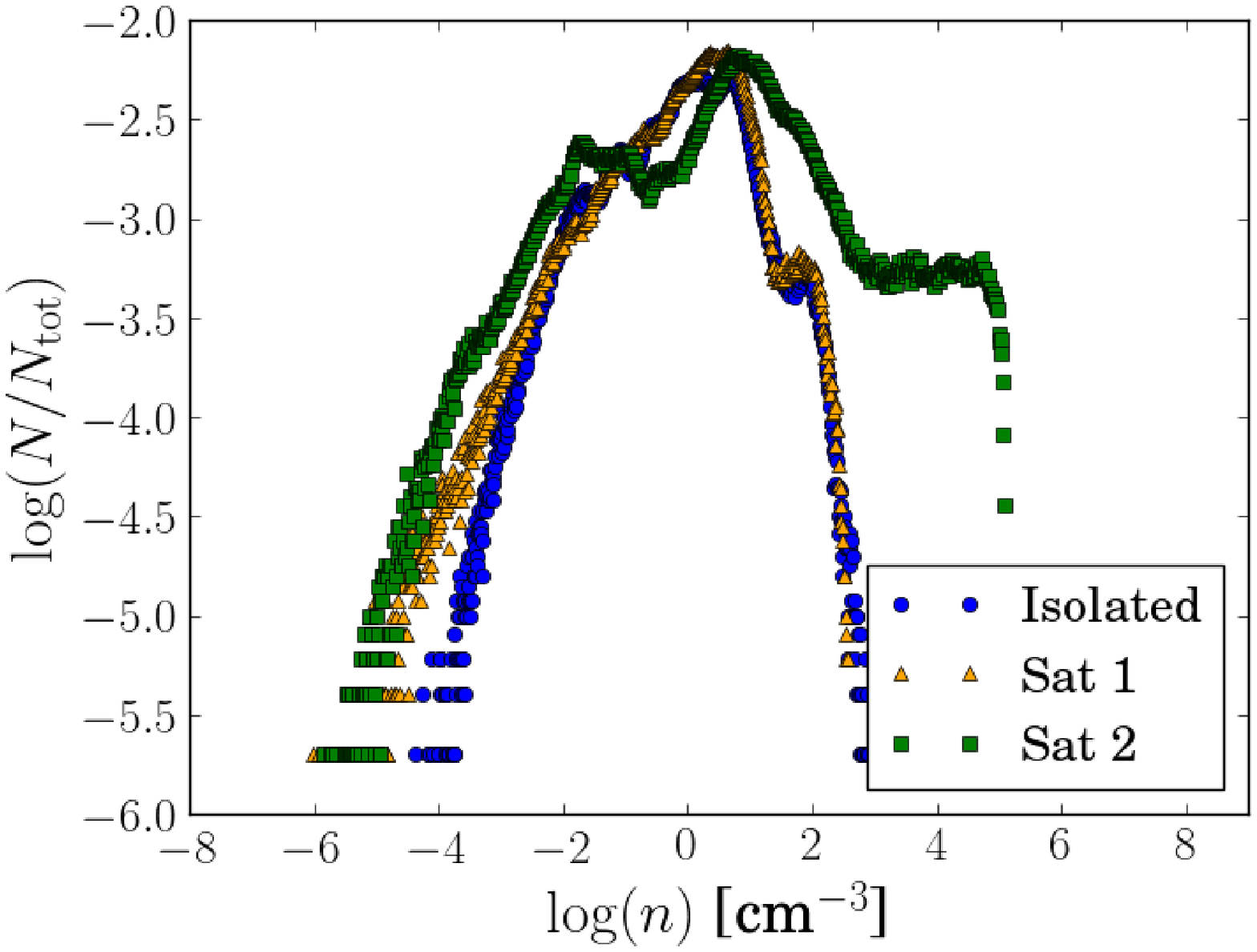}
		\caption{Number density distributions for the simulation of Model A with Orbit 2 ($R_i = 6R_D,\, i = 30^{\circ}$) in a prograde sense. This was obtained at $t \approx 470$ Myr since the beginning of the simulation. The change takes longer because of the wider orbit of the satellite.}
		\label{fig:dens-dist-mod-B-O2-Pro-t2}
	\end{center}
\end{figure}

For the simulation of Model A and prograde Orbit 2, a longer time passes before the distribution starts to change in comparison to the case with Orbit 1, which can be attributed to the longer time it takes the satellite to reach the disc. The distributions at $t \approx 282$ Myr are shown in the middle left panel of Figure \ref{fig:dens-dist-mod-A}. For the simulation with Satellite 2, the density distribution shows gas in higher densities at later times. Figure \ref{fig:dens-dist-mod-B-O2-Pro-t2} plots the distribution at $t \approx 470$ Myr. The simulation with Satellite 1 does not produce any noticeable changes in the density distribution at this time, even though it has triggered some inward flows as shown in \S \ref{subsubsec:results-flows-modelA}. For the same model and orbit but in a retrograde sense, both the volume distributions are not significantly different at $t = 282$ Myr (see the middle right panel of Figure \ref{fig:dens-dist-mod-A}).

For the simulation with prograde Orbit 3 and Model A, a significant change in the density distribution is seen beginning at $t \approx 188$ Myr for the simulations with both satellites. The change is larger for Satellite 2. The high-density tail of the distribution continues growing with time, with the simulation with Satellite 2 producing slightly higher densities than the one with Satellite 1 (shown in the bottom left panel of Figure \ref{fig:dens-dist-mod-A} at $t = 282$ Myr). However, the distribution doesn't seem to be strongly sensitive to the satellite's mass in this case. In the retrograde version of this simulation, similar results are obtained (shown in the bottom right panel of Figure \ref{fig:dens-dist-mod-A} at $t = 282$ Myr).

% Gas density distributions compared at t = 60 u_t

% Density distribution at a later time for Orbit 2 Prograde and Model A.

The simulations with the barred model show qualitatively similar results to those of Model A, which we briefly describe below. In the simulation of Model B and prograde Orbit 1, the density distribution shows some slight changes with respect to that of isolated evolution, but it is not significantly different. For the retrograde version of this simulation, the density distribution does not develop any large differences with respect to isolated evolution. In the simulation of Model B and prograde Orbit 2, the density distribution does not show marked differences with respect to isolated evolution, although a small peak in the distribution around $10^6$~cm$^{-3}$ starts to grow with time. In the retrograde version of this simulation, the differences in distributions between the isolated and perturbed models are not significant. In the case of the simulation of Model B and Orbit 3, the density distribution also does not show strong differences with respect to isolated evolution. As with the case of Orbit 2, a peak in the distribution develops with time around $10^6$~cm$^{-3}$.

In both models, $\approx 80\%$ of the gas has densities higher than $10^{-1}$~cm$^{-3}$, which in some simulations has been taken as a threshold for star formation activity (e.~g.~\citealt{SchayeandDallaVecchia2008,Starkenburgetal2016}). We see that the density distributions tend to change for gas at higher densities than this value. When we use $10^1$~cm$^{-3}$ as reference, $\approx 10\%$ of the gas has higher densities for Model A, and $\approx 30-40\%$ for Model B in isolation. In the interacting cases, the amount of gas denser than $10^1$~cm$^{-3}$ increases, particularly for the prograde inclined and the coplanar orbits in both senses. For model A and Prograde Orbit 1, the amount of gas increases by about a factor of 3 (see Figure \ref{fig:dens-dist-mod-A}) and for Prograde Orbit 2, it increases by a factor of $\approx 4$ (Figure \ref{fig:dens-dist-mod-B-O2-Pro-t2}). For the coplanar orbits, both prograde and retrograde, it also increases by a factor of $4$. These results are for Satellite 2. The amounts for Satellte 1 are slightly lower, but still significant. For the simulations of Model B, the highest change is for Prograde Orbit 2 and Satellite 2, and amounts to $\approx 30\%$. For the coplanar orbits, the change can be up to $\approx 50\%$. The difference with respect to the values for Model A is mainly due to the fact that the spiral arms and bar in Model B drive higher gas densities in isolation compared to Model A. However, we note that these density distributions should be interpreted carefully due to our assumption of an isothermal gas. We are not including mechanisms that may drive the formation of the different ISM phases, which would change the overall density distribution in isolated evolution.

The overall tendency is that the dense gas tends to form in the spiral arms and in some dense clumps that are eventually formed in the arms. For Model B, it is possible to compare the peak gas densities in the arms in the perturbed galaxies with those of the isolated model. In general, we don't find any clear tendency. In the inner regions, the peak densities are not significantly different to those in the isolated model. However, in the outer galaxy, the peak densities in the perturbed galaxies are more important as the amplitude of the arms of the isolated model is much lower in those regions. Some simulations of isolated galaxies do show that the peak arm density tends to decrease with radius (e.~g.~\citealt{Ramon-Fox2019}).

%%%%%%%%%%%%%%%%%%%%%%%%%%%%%%%%%%%%%%%%%%%%%%%%%%
%%%%%%%%%%%%%%%%%%%%%%%%%%%%%%%%%%%%%%%%%%%%%%%%%%
%%%%%%%%%%%%%%%%%%%%%%%%%%%%%%%%%%%%%%%%%%%%%%%%%%
\section{Discussion}
\label{sec:discussion}

%%%%%%%%%%%%%%%%%%%%%%%%%%%%%%%%%%%%%%%%%%%%%%%%%%
%%%%%%%%%%%%%%%%%%%%%%%%%%%%%%%%%%%%%%%%%%%%%%%%%%
\subsection{Galaxy Morphology}

The results in \S \ref{sec:results} show that the satellites in prograde orbits produce evident morphological features in both the gas and stellar components. In our simulations, these encounters generate a spiral pattern resembling a grand-design galaxy. Satellites in these orbits are expected to produce higher damage to the primary galaxy due to the low relative velocity of the perturber with respect to disc's rotation. The slow relative motion allows the satellite to significantly alter the orbits of the gas and stars in the region where it passes through the disc. Our simulations also show that very low-mass satellites in some retrograde orbits can produce noticeable morphology changes, but this becomes apparent after several disc crossings. However, some of our simulations consider circular orbits, which results in a periodic perturbation that enhances the effect of the retrograde interaction.

The morphology of the prograde encounters found here is qualitatively similar to those obtained in previous works that have used more massive satellites. Orbit 2 ($R_i = 6R_D$, $i = 30^{\circ}$) was first used in \citet{MihosHernquist94} and \citet{HernquistMihos95}. The morphology of our simulated galaxies is qualitatively similar to that in these works. The spiral arm morphology in the stellar component of our simulations is similar to those of other studies of infalling satellites such as \citet{VelazquezWhite1999}, \citet{Kazantzidis08}, and \citet{Kazantzidis09}. The present work considers satellites with a mass ratio between $\approx$6:1000 and $\approx$3:265 whereas \citet{VelazquezWhite1999} use satellites with mass ratios between $\approx$1:125 and $\approx$3:200. \citet{Kazantzidis08} and \citet{Kazantzidis09}, use satellites with a mass ratio in the range between $\approx$1:100 and $\approx$1:38. Although the latter used orbits derived from cosmological simulations. Similarly, the simulations by \citet{HuandSijacki2018}, which use satellites with properties extracted from cosmological simulations in the same mass range as our satellites, show similar morphological features. However, neither \citet{Kazantzidis09} nor \citet{HuandSijacki2018} included hydrodynamics.

\citet{Chakrabarti11} studied the tidal effect of satellites with mass ratios in the range between 1:100 and 1:3 to obtain models that fit the observed spatial gas distributions in galaxies such as M51 and NGC 1512. Additionally, they explore if the mass of an interacting satellite could be inferred from the observed morphology. The morphology in our simulations is similar to those obtained in the simulations of \citet{Chakrabarti11}, although they only considered prograde orbits. On the other hand, our simulations with retrograde orbits show that the interaction is capable of breaking the disc's axisymmetry, but the strength of the effect has some dependence on the satellite's mass and its orbital inclination. More recently, \citet{Kyziropoulosetal2016} also explored the effect of interactions in a similar mass range also show the formation of two-armed as well as asymmetric discs, although they did not focus on the large-scale gas flows.

Observations have shown that many galaxies present some \emph{lopsidedness} or asymmetries in their gaseous and stellar components; see for example \citet{BealeDavies1969,Baldwinetal1980,RichterSancisi94,RixZaritsky95,ZaritskyRix97,Bournaudetal05,JogCombes09,Zaritskyetal2013,Boketal2019}. \cite{Matthewsetal1998} found that a fraction of $77 \%$ of a sample of late-type galaxies have lopsided HI profiles, which they attribute to lopsided distributions. Simulations by \citet{Bournaudetal05} indicate that mergers with a mass ratio $\approx$1:10 produce lopsidedness with $m = 1$ and $m = 2$ modes. The present work shows that even less massive satellites can produce these effects.

On other hand, our satellites are small enough that they will not lead, for example, to the formation of an spheroidal component upon merging (e.~g.~\citealt{Kodaetal2009}). Similarly, not much vertical heating is expected given the very small mass of our satellite (e.~g.~\citealt{TothandOstriker1992,MoetazedianandJust2016}). A quantitative assessment of such effects on the disc is out of the scope of the present paper. However, hydrodynamical effects in the disc due to the bombardment of multiple satellites with different orbits and a spectrum of masses is clearly of importance in the evolution of disc galaxies (e.~g.~\citealt{DOnghiaetal2016}).

%%%%%%%%%%%%%%%%%%%%%%%%%%%%%%%%%%%%%%%%%%%%%%%%%%
%%%%%%%%%%%%%%%%%%%%%%%%%%%%%%%%%%%%%%%%%%%%%%%%%%
\subsection{Gas Flows}

The results in \S \ref{subsec:mass-distribution} show that satellites with $\mathcal{R} \approx 1:1000 - 1:100$ produce significant flows in the host's disc, but this is sensitive to the satellite's orbital parameters and mass.

%%%%%%%%%%%%%%%%%%%%%%%%%%%%%%%%%%%%%%%%%%%%%%%%%%
\subsubsection{Effect of Orbital Parameters and Satellite Mass}

In our simulations, all prograde orbits produce significant gas flows except for Orbit 1with the barred model. The satellite in Orbit 2 produces the strongest effects and shows the most sensitivity to the satellite's mass. The results for Orbit 1 may be attributed to the fact that the satellite passes through inner regions with higher density and gravitationally more bound, thus being harder to disturb. This effect and the orbit's prograde sense favour a quick disruption of the satellite, which reduces the intensity of the perturbation. For Orbit 2, the satellite first interacts with the outer, less dense, and less bound regions of the galaxy. Additionally, a stronger spiral pattern is formed in both the stellar and gas discs, which produce torques that alter the gas motions.

In the case of a prograde coplanar orbit (Orbit 3), both models show that inward flows were induced and the gas distribution is more noticeably changed in Model A than Model B. The effect of this orbit appears to be less sensitive to the satellite's mass. On the other hand, the retrograde encounter following Orbit 3 also produces inward flows, without a clear dependence on the satellite's mass. The coplanar orbit simulations with both galaxy models show similar final gas mass distributions.

%%%%%%%%%%%%%%%%%%%%%%%%%%%%%%%%%%%%%%%%%%%%%%%%%%
\subsubsection{Comparison Between Non-Barred and Barred Model}

Model A does not form a bar or strong spiral structure, which allows to isolate the effects of the interacting satellite from those of other galactic structure. The simulations with Model A show that low-mass satellites can still produce significant gas inflows in the host galaxy. With the most massive satellite in Orbit 2, approximately $35$\% of the galaxy's gas passed through $R = R_D$ in a timescale of $t \approx 2.5\tau$. This corresponds to $\approx 1.6 \times 10^{9} \msun$ in $\approx 600$ Myr. The interacting satellite induces the formation of non-axisymmetric features in the gaseous and stellar components which may redistribute the angular momentum of the gas and produce inflows. As the gas concentrates in the spiral arms, it can loose orbital energy through dissipation and shocks. It is noted that although an artificial viscosity term is introduced in SPH to treat shocks, these may not be adequately represented at the resolution of our simulations. The present work focuses on the large-scale gas dynamics, which are adequately represented. \citet{KimKim2014} show that spiral structure can have an important effect in driving gas flows in galaxies. In this sense, our simulations with the most prominent spiral features tend to show the highest inflows. 

An interesting result is that of Orbit 3, which shows that some particular retrograde orbits produce significant flows. In this case, approximately $20$\% of the gas entered the region where $R < R_D$. This orbit also triggers the formation of spiral arms in the disc. The constant presence of the satellite in the disc's plane introduces an additional perturbation to the potential that disturbs the gas motions. Although such a finely tuned coplanar orbit may be rather unlikely, the simulation shows that the cumulative effect of a retrograde orbit is not negligible. The difference between the prograde and retrograde simulations of Orbit 3 are noticeable in the flows passing by $R = 0.28R_D$. For the prograde one, about $30$\% of the gas moves to $R < 0.28R_D$ whereas in the retrograde one, the fraction is slightly more than $20$\%. The simulations with Model A show that satellites with small masses compared to the host still have a significant effect on the gas dynamics of the host.

The similarity of the final gas distributions of Models A and B in both the prograde and retrograde orbits supports the fact that the satellite's perturbation is inducing flows in the host. However, in the simulations with prograde Orbit 1, the satellite disturbs the gas in the inner parts of the galaxy. In Model B, the interaction occurs when the bar has already redistributed material in the inner regions, so a certain fraction of the galaxy's gas is no longer available to be further displaced by the satellite. This can explain the lack of significant changes in Model B with respect to isolated evolution (top left panel in Figure \ref{fig:mass-distributions-model-A}) On the other hand, the simulations with Orbit 2 do show that the the interaction is contributing with additional gas flows. The simulations of Model B in isolated evolution show that the mass fraction at $R \approx 2.28R_D$ stays at $\approx 60 \%$ during 8 orbital periods, and the gas disc extends up to about $(5 - 6)R_D$. Therefore, about $40 \%$ of the galaxy's gas is yet available to be redistributed by the interaction.

The simulations of Model B and Orbit 3 show that both the prograde and retrograde orbits produce inflows that are somewhat higher than those in isolated evolution. The dependence of the final gas distribution on the satellite's mass is weak. An interesting result is that the retrograde simulation with Orbit 3 also produces noticeable gas inflows in Model B. Slightly less than $10\%$ of the galaxy's gas moves to within $R = R_D$. However, at $R = 0.28R_D$ the difference with respect to isolated evolution is not significant. Although a stellar bar's effect depends on its strength and size, it can drive continuous gas flows (e. g. \citealt{Athanassoula94,ReganTeuben2004,KimSeoKim2012}). These results suggest that minor interactions may become more important in triggering flows when they disturb the outer parts of a barred galaxy.

%%%%%%%%%%%%%%%%%%%%%%%%%%%%%%%%%%%%%%%%%%%%%%%%%%
\subsubsection{Comparison with Previous Numerical Work}

\citet{MihosHernquist94} and \citet{HernquistMihos95} explored the role of minor mergers as a mechanism for inducing activity in galaxies. They used a model galaxy with an isothermal dark matter halo, stellar and gas discs. They alternated between models with bulge and no bulge. Their model assumed $Q(R_0) = 1.5$ at the solar radius. These authors indicate that no significant gas flows are observed in isolated evolution. With the interaction, they obtain that $\approx 50 \%$ of the gas moves to $R < R_D$ in a bulgeless galaxy. This quantity decreases to $\approx 35 \%$ when the mass of the bulge is $1/3$ of the disc's mass. However, they do not comment in detail about the of the presence of a bar in their models, making a comparison with their results rather difficult.

Our work uses a NFW dark matter halo instead of an isothermal one, and also includes stellar and gaseous discs. A bulge is included in all simulations, with $M_b \approx 1/5M_d$. We also use a model galaxy with a rather high Toomre parameter ($Q = 3$), not explored in previous works. \citet{HernquistMihos95} use a satellite with a Hernquist profile with $M_s = 0.1M_D$, which corresponds to $\mathcal{R} \approx 1:72$. Our work, uses satellites with a Plummer profile, with $M_s \approx 0.13M_D$ and $M_s \approx 0.3M_D$, corresponding to approximate mass ratios of 6:1000 and 3:265, respectively.

The less massive satellite used in the present work has a much smaller mass ratio than that in \citet{HernquistMihos95}. The bulge mass of the galaxy models used in the present work falls between the masses used by \citet{HernquistMihos95}. The simulation with Model A has a slightly higher inflow at $R = R_D$ than that of \citet{HernquistMihos95}. The cumulative gas mass fraction in our work is similar to that of the less massive bulge in \citet{HernquistMihos95}. The results of Model B also show values slightly higher than those of \citet{HernquistMihos95}. The final integrated mass functions of Orbit 2 with the less massive satellite are qualitatively similar to those of \citet{HernquistMihos95}, and that of Model B shares more similarities in the behaviour as a function of radius. However, our results show a higher mass fraction at $R = 2R_D$ than in \citet{HernquistMihos95}. An important difference between \citet{HernquistMihos95} and this work is that we use a $\approx 3.9$ times more massive dark matter halo with a NFW profile. A different profile can affect the disc's dynamics as the possible orbits in the central region may change, and a more massive halo implies a deeper potential well. However, the similar results suggest that a difference in the halo profile may not be of strong importance for gas motions at galactic disc scales, in comparison to a difference in the halo's mass. However, the halo's inner profile slope may play a more noticeable role in the gas dynamics of the galaxy's inner region.

Our work has also explored a set of retrograde orbits. Such cases are relatively unexplored in the literature because less damage is expected from satellites in these orbits. In the case of inclined orbits such as Orbit 1 and 2, no significant inflows are obtained, which is expected from the high relative velocity with respect to the disc. However, a coplanar retrograde orbit such as Orbit 3 produces significant gas flows regardless of the presence of a bar. In the case of Model A, the final gas mass fraction within $R = R_D$ increases by $\approx 60$\% with respect to isolated evolution, with no significant dependence on the satellite's mass. In the case of Model B, the increase is about $12$\%, also with no clear difference with respect to the satellite's mass. The movement of the small satellite along the plane of the galaxy produces morphological features that break the disc's symmetry. The perturbation of the gravitational potential induced by these density features as well as the continuous presence of the satellite maintain a certain inflow rate. Although such a fine-tuned orbit may be unlikely for an infalling satellite, it is an example of a retrograde orbit that produces noticeable effects in the dynamics of the primary galaxy, situation that has not been explored or reported earlier in similar works.

\citet{BekkiChiba2006} and \citet{Kannanetal2012} have also studied mergers in the range of 1:1000 to study the formation of density enhancements and ``holes'' in the gas density distribution assuming vertically falling orbits. We assume a different set of orbital parameters, so we cannot directly compare our results. However, we do see that the passage of the satellite does produce some density enhancements during its passage, particularly for Orbit 1. An example of these features for Orbit 2 is shown in Figure \ref{fig:morph-orbit2-pro}. However, our simulations produce strong arms, which drive most of the surface density enhancements with respect to the isolated models (see Figures \ref{fig:morph-orbit2-pro} and \ref{fig:morph-orbit1-ret}).

Our morphological results agree with those of \citet{Starkenburgetal2016}, who study simulations of interactions with mass ratios in the range of 1:5 - 1:20. They find that the interaction drives significant asymmetries in the gas and stellar distributions. Our results complement this work in the sense that even smaller satellites can still produce important effects in the morphology. In terms of triggered spiral arms, our results also agree qualitatively with those of the simulations of \citet{Chakrabarti11} and \citet{Shahetal2019}, who used satellites in the range of  $\mathcal{R} \approx 1:100 - 1:1000$.

\citet{Pettittetal2016} also explored the effect of minor interactions with satellites in the range $\mathcal{R} \approx 19:100 - 5:1000$, which is similar to that of our work. However, they assumed interactions following parabolic orbits. They find that satellites with $M > 10^9 \msun$ can trigger spiral structure in the disc. They also find that a satellite with $M = 10^{10} \msun$ can drive some flows on the primary, which agrees to some extent with our results. They find lower flow rates because their orbit does not have such a periodic perturbation as the circular orbit assumed in our work. Their results also show that the triggered flows are sensitive to the satellite's mass and orbital parameters.

%%%%%%%%%%%%%%%%%%%%%%%%%%%%%%%%%%%%%%%%%%%%%%%%%%
%%%%%%%%%%%%%%%%%%%%%%%%%%%%%%%%%%%%%%%%%%%%%%%%%%
\subsection{Astronomical Applications}

We now briefly discuss two astronomical implications of our work.

%%%%%%%%%%%%%%%%%%%%%%%%%%%%%%%%%%%%%%%%%%%%%%%%%%
\subsubsection{Lopsidedness and Asymmetries in Discs}

Several isolated galaxies (e.~g.~\citealt{Karachentsev1972}) show signs of disturbances like lopsidedness. However, there is no clear evidence of the presence of similar neighbouring companions. Due to the low surface brightness of dwarf galaxies, it is difficult to determine observationally if a population of these objects exists around larger host galaxies \citep{Bullocketal2010b}. According to the standard galaxy formation scenario (e.~g. \citealt{MoVanDenBoschWhite2010}), a significant population of satellites is expected around galaxies, which is reproduced in cosmological simulations (e.~g. \citealt{Klypinetal1999}; see \citet{Bullock2010} for a review). This implies that galactic discs will have experienced of the order of thousands of minor interactions leading to impulsive shocks and disc heating \citep{Mooreetal1999}. 

Other simulations have explored the effect of minor mergers with mass ratios in the range of $\sim$1:10 and their effects in the primary's morphology (e.~g.~\citealt{Dobbsetal2010,Purcelletal2010,Purcelletal2011,Quetal2011}), while others those in the range $1:100 - 1:1000$ (e.~g.~\citealt{Chakrabarti11,Shahetal2019}). Our work shows and agrees with the latter ones in the sense that even satellite galaxies, smaller than roughly the size of the Small Magellanic Cloud, can produce lopsided density distributions in galaxies similar to the Milky Way. Furthermore, their effects may be more noticeable in HI gas maps (e.~g.~\citealt{ChangandChakrabarti2011,Lipnickyetal2018}). Considering that Milky Way-sized haloes may have had an active accretion history, these satellites can still have an impact on these systems.

Observations by \citet{Wagner-Kaiseretal2014} show an interesting $\approx 3$ kpc large ``void'' in the HI distribution of NGC 247, which the authors attribute to an interaction with a dark subhalo. Our simulations show that the passage of the satellite does induce some local density features. However, it is interesting to note that the simulation with retrograde Orbit 1 and Model A develops an asymmetry in the gas distribution and a low surface density feature in the lower galaxy that appears qualitatively similar. Nevertheless, this comparison should be taken carefully considering that our models do not include the effects of star formation and feedback.

Our results support the studies of \citet{RudnickandRix1998} and \citet{Rudnicketal2000} in the sense that minor interactions can explain asymmetries in discs. Our density distributions suggest that in some cases the star formation activity could potentially increase due to small interactions which have implications for the study of \citet{Edmanetal2012}, which show that a fraction of galaxies with an enhanced star formation rate have a companion in a sample from SDSS. However, a more solid result would require redoing our simulations using a more realistic model of the ISM, star formation activity, and feedback, which would allow a more meaningful comparison.

%%%%%%%%%%%%%%%%%%%%%%%%%%%%%%%%%%%%%%%%%%%%%%%%%%
\subsubsection{Active Galactic Nuclei}

Our results also have implications in the triggering of nuclear activity in galaxies. In the context of Active Galactic Nuclei (or AGNs), it is widely accepted that mergers are a triggering mechanism (e. g \citealt{Taniguchi1999,AlexanderHickox2012}), but it is not clear from an observational standpoint to what extent minor mergers and perturbations of small satellites can have an impact. Additionally, there is no clear correlation between the presence of an AGN and a bar in the host galaxy (for a review, see \citealt{BeckmannShrader2012}). A recent study using SDSS data by \citet{Gallowayetal2015} finds that barred galaxies hosting an AGN do not show a higher accretion rate than unbarred AGN hosts. This suggests that other mechanisms are relevant in driving flows that fuel the AGN. \citet{Chang2008} has explored the potential of minor mergers in feeding the central region and triggering nuclear activity, but only using approximate analytical models, and \citet{Altamiranoetal2016} their probable relevance in explaining some large-scale clustering behaviour of AGNs. This work has explored this point using more realistic galaxy models. The simulations with Model A show that a small external disturbance can still play an important role in driving flows in a non-barred galaxy.

For a galaxy with a weak stellar bar or none at all, these small interactions can be a viable mechanism for driving important gas flows to the central regions. However, other effects such as star formation and feedback should also be included for a more complete modelling. In the case of barred galaxies, the simulations with Model B show that these minor satellites can have an important effect as long as the bar has not displaced significant amounts of gas. Nevertheless, gas from the galaxy's outer regions may still be redistributed by the interaction. 

Most of the prograde orbits seem favourable for driving gas to the central regions ($R < 0.28R_D$). When scaled to $R_D = 3.5$ kpc, as for the Milky Way galaxy, this means that the displaced gas is reaching the inner kpc of the host. When the satellite disturbs the outer regions, the gas can take up to $\approx \tau$ to reach the central regions. This result is relevant in the sense that if these mergers leave transient morphological features, they may disappear before any nuclear activity is triggered. In the case of the retrograde circular orbits, no significant flows were obtained, but noticeable morphological effects were produced, specially with a more massive satellite. Although these orbits may be unimportant for nuclear activity, they may be relevant in triggering some density enhancements that may lead to localised star formation activity.

We find in some of our simulations that a gas fraction comparable to $\approx 50 \%$ of the galaxy's gas ($\approx 2.2 \times 10^9 \msun$) ends at the region where $R < R_D$. In the most extreme case (Orbit 2), approximately $35\%$ of the gas moves to this region from the outer parts of the disc during the simulation, which corresponds to a mass of $\approx 1.5 \times 10^9 \msun$. In our simulations, the gas is driven to the galaxy's central regions in a timescale of at least $\approx 240$ Myr. 

In simulations of gas flows in interacting galaxies, \citet{HopkinsQuataert2010} find an accretion rate of $\dot{M} \sim 10  \msunperyear$ at scales of $0.1$ pc. The analysis of the NUGA sample of nearby active spiral galaxies by \citet{Haanetal09} finds accretion rates in the range of $0.01 - 50 \msunperyear$ at scales of from 1 kpc to 10 pc with typical flow timescales of $\sim 10^2$ Myr. The simulations of \citet{Montuori2010} and \citet{Pettittetal2016} find typical timescales for inflows also in the order of $10^2$ Myr. These are similar to the timescale we find in our work. Assuming a mass flow rate of  $\sim10 \msunperyear$ (e. g. \citealt{HopkinsQuataert2010}) in a timescale of $240$ Myr, the amount of accreted gas is $2.4 \times 10^9 \msun$, which is comparable to the gas fraction that moves to within $R = R_D$ found in some of our simulations. It is interesting to note that this timescale is consistent with the duration of starbursts triggered by mergers found by \citet{DiMatteoetal2008} in major interactions. Our results may also be relevant in the context of driving star formation activity. According to \citet{Kennicutt98}, displacing a fraction of up to $50$\% of the gas of the galaxy may be needed to fuel the star formation observed in some active galaxies.

Studies such as \citet{Willettetal2015} and \citet{Satyapaletal2014} of observational samples of galaxies with AGN suggest that mergers are a relevant mechanism in triggering nuclear activity. Our simulations support these results. However, we note that our interpretation may be limited by the simplicity of our model. Certainly, it would be important to repeat our simulations with more detailed ISM and star formation physics, as well as including some model of the central black hole.

%%%%%%%%%%%%%%%%%%%%%%%%%%%%%%%%%%%%%%%%%%%%%%%%%%
%%%%%%%%%%%%%%%%%%%%%%%%%%%%%%%%%%%%%%%%%%%%%%%%%%
%%%%%%%%%%%%%%%%%%%%%%%%%%%%%%%%%%%%%%%%%%%%%%%%%%
\section{Conclusions}
\label{sec:conclusions}

We explore the role of small low-mass satellites, with total mass ratio ${\cal R}\approx 1:1000-1:100$, in driving large amounts of gas to the central disc of a galaxy resembling the Milky Way. We considered both a barred and non-barred model. Our simulations show that these interactions produce noticeable morphological features in the primary's disc. Prograde orbits produced grand-design spiral arms whereas retrograde interactions produced asymmetries or lopsidedness. 

The simulations also show that these small interactions trigger significant gas flows to the galaxy's centre. The most extreme case shows that a fraction of $\approx 60 \%$ of the total gas in the disc falls within $1R_D$ of a Milky Way sized galaxy after the interaction and merging of the satellite. The simulation of an interaction with a non-barred galaxy shows that the satellite perturbation produces a significant change in the gas distribution. Our simulations show that an important amount of gas can reach the central regions of a disc galaxy due to the interaction with a small satellite.

For future work, it would be interesting to perform simulations using satellite orbital parameters derived from cosmological simulations. Our model is also limited by assuming an isothermal gaseous component for the ISM, lacking other physical processes such as star formation and feedback.

%%%%%%%%%%%%%%%%%%%%%%%%%%%%%%%%%%%%%%%%%%%%%%%%%%
%%%%%%%%%%%%%%%%%%%%%%%%%%%%%%%%%%%%%%%%%%%%%%%%%%
%%%%%%%%%%%%%%%%%%%%%%%%%%%%%%%%%%%%%%%%%%%%%%%%%%
\section*{Acknowledgments}
 
We wish to thank the anonymous referee for the valuable comments provided. This research was funded by UNAM-PAPIIT and CONACyT Research Projects IN109710, IN104113 and 179662, respectively. FGR-F acknowledges support from a CONACyT postgraduate scholarship during the first part of this project at UNAM, support from the European Research Council ECOGAL project (agreement 291227) at the University of St Andrews, and the Hyperstars project, which is supported by R\'egion Paris-I\^le-de-France (DIM-ACAV) and the MASTODONS initiative of the CNRS. We thank Lars Hernquist for helpful comments at the beginning of this project a few years ago. We also warmly thank Edilberto S\'anchez for his technical support at the HPC facilities at IA-UNAM, Ensenada. This work has made use of NASA's Astrophysics Data System.

%%%%%%%%%%%%%%%%%%%%%%%%%%%
\bibliographystyle{mnras} %{mn2e}
\bibliography{biblio}

\begin{thebibliography}{}
\makeatletter
\relax
\def\mn@urlcharsother{\let\do\@makeother \do\$\do\&\do\#\do\^\do\_\do\%\do\~}
\def\mn@doi{\begingroup\mn@urlcharsother \@ifnextchar [ {\mn@doi@}
  {\mn@doi@[]}}
\def\mn@doi@[#1]#2{\def\@tempa{#1}\ifx\@tempa\@empty \href
  {http://dx.doi.org/#2} {doi:#2}\else \href {http://dx.doi.org/#2} {#1}\fi
  \endgroup}
\def\mn@eprint#1#2{\mn@eprint@#1:#2::\@nil}
\def\mn@eprint@arXiv#1{\href {http://arxiv.org/abs/#1} {{\tt arXiv:#1}}}
\def\mn@eprint@dblp#1{\href {http://dblp.uni-trier.de/rec/bibtex/#1.xml}
  {dblp:#1}}
\def\mn@eprint@#1:#2:#3:#4\@nil{\def\@tempa {#1}\def\@tempb {#2}\def\@tempc
  {#3}\ifx \@tempc \@empty \let \@tempc \@tempb \let \@tempb \@tempa \fi \ifx
  \@tempb \@empty \def\@tempb {arXiv}\fi \@ifundefined
  {mn@eprint@\@tempb}{\@tempb:\@tempc}{\expandafter \expandafter \csname
  mn@eprint@\@tempb\endcsname \expandafter{\@tempc}}}

\bibitem[\protect\citeauthoryear{Alexander \& Hickox}{Alexander \&
  Hickox}{2012}]{AlexanderHickox2012}
Alexander D.~M.,  Hickox R.~C.,  2012, New Astronomy Reviews, 407, 93

\bibitem[\protect\citeauthoryear{{Altamirano-D{\'e}vora}, {Miyaji}, {Aceves},
  {Castro}, {Ca{\~n}as}  \& {Tamayo}}{{Altamirano-D{\'e}vora}
  et~al.}{2016}]{Altamiranoetal2016}
{Altamirano-D{\'e}vora} L.,  {Miyaji} T.,  {Aceves} H.,  {Castro} A.,
  {Ca{\~n}as} R.,   {Tamayo} F.,  2016, \rmxaa, \href
  {https://ui.adsabs.harvard.edu/abs/2016RMxAA..52...11A} {52, 11}

\bibitem[\protect\citeauthoryear{Athanassoula}{Athanassoula}{1994}]{Athanassoula94}
Athanassoula L.,  1994, in Mass-Transfer Induced Activity in Galaxies.
  Cambridge University Press, p.~143

\bibitem[\protect\citeauthoryear{Baldwin, Lynden-Bell  \& Sancisi}{Baldwin
  et~al.}{1979}]{Baldwinetal1980}
Baldwin J.~E.,  Lynden-Bell D.,   Sancisi R.,  1979, MNRAS, 193, 313

\bibitem[\protect\citeauthoryear{{Barber}, {Starkenburg}, {Navarro},
  {McConnachie}  \& {Fattahi}}{{Barber} et~al.}{2014}]{Barberetal2014}
{Barber} C.,  {Starkenburg} E.,  {Navarro} J.~F.,  {McConnachie} A.~W.,
  {Fattahi} A.,  2014, MNRAS, 437, 959

\bibitem[\protect\citeauthoryear{{Barnes}}{{Barnes}}{2002}]{Barnes2002}
{Barnes} J.~E.,  2002, \mn@doi [\mnras] {10.1046/j.1365-8711.2002.05335.x},
  \href {http://adsabs.harvard.edu/abs/2002MNRAS.333..481B} {333, 481}

\bibitem[\protect\citeauthoryear{Barnes \& Hernquist}{Barnes \&
  Hernquist}{1991}]{BarnesandHernquist1991}
Barnes J.,  Hernquist L.,  1991, ApJ, 370, L65

\bibitem[\protect\citeauthoryear{{Barnes} \& {Hernquist}}{{Barnes} \&
  {Hernquist}}{1996}]{BarnesandHernquist1996}
{Barnes} J.~E.,  {Hernquist} L.,  1996, \mn@doi [\apj] {10.1086/177957}, \href
  {http://adsabs.harvard.edu/abs/1996ApJ...471..115B} {471, 115}

\bibitem[\protect\citeauthoryear{Barnes \& Hut}{Barnes \&
  Hut}{1986}]{BarnesHut86}
Barnes J.,  Hut P.,  1986, Nature, 324, 446

\bibitem[\protect\citeauthoryear{Beale \& Davies}{Beale \&
  Davies}{1969}]{BealeDavies1969}
Beale J.~S.,  Davies R.~D.,  1969, Nature, 193, 313

\bibitem[\protect\citeauthoryear{Beckmann \& Shrader}{Beckmann \&
  Shrader}{2012}]{BeckmannShrader2012}
Beckmann V.,  Shrader C.,  2012, {Actice Galactic Nuclei}.
Wiley-VCH

\bibitem[\protect\citeauthoryear{Bekki \& Chiba}{Bekki \&
  Chiba}{2006}]{BekkiChiba2006}
Bekki K.,  Chiba M.,  2006, ApJ, 637, L97

\bibitem[\protect\citeauthoryear{Benson}{Benson}{2005}]{Benson2005}
Benson A.~J.,  2005, MNRAS, 634, 551

\bibitem[\protect\citeauthoryear{{Blumenthal} \& {Barnes}}{{Blumenthal} \&
  {Barnes}}{2018}]{BlumenthalandBarnes2018}
{Blumenthal} K.~A.,  {Barnes} J.~E.,  2018, \mn@doi [\mnras]
  {10.1093/mnras/sty1605}, \href
  {https://ui.adsabs.harvard.edu/abs/2018MNRAS.479.3952B} {479, 3952}

\bibitem[\protect\citeauthoryear{{Bok}, {Blyth}, {Gilbank}  \& {Elson}}{{Bok}
  et~al.}{2019}]{Boketal2019}
{Bok} J.,  {Blyth} S.~L.,  {Gilbank} D.~G.,   {Elson} E.~C.,  2019, \mn@doi
  [\mnras] {10.1093/mnras/sty3448}, \href
  {https://ui.adsabs.harvard.edu/abs/2019MNRAS.484..582B} {484, 582}

\bibitem[\protect\citeauthoryear{Bournaud, Combes, Jog  \& Puerari}{Bournaud
  et~al.}{2005}]{Bournaudetal05}
Bournaud F.,  Combes F.,  Jog C.~J.,   Puerari I.,  2005, A\&A, 438, 507

\bibitem[\protect\citeauthoryear{{Bullock}}{{Bullock}}{2010}]{Bullock2010}
{Bullock} J.~S.,  2010, preprint, p.~1043 (\mn@eprint {} {1009.4505})

\bibitem[\protect\citeauthoryear{{Bullock}, {Stewart}, {Kaplinghat}, {Tollerud}
   \& {Wolf}}{{Bullock} et~al.}{2010}]{Bullocketal2010b}
{Bullock} J.~S.,  {Stewart} K.~R.,  {Kaplinghat} M.,  {Tollerud} E.~J.,
  {Wolf} J.,  2010, ApJ, 717, 1043

\bibitem[\protect\citeauthoryear{{Bustamante}, {Sparre}, {Springel}  \&
  {Grand}}{{Bustamante} et~al.}{2018}]{Bustamante_etal2018}
{Bustamante} S.,  {Sparre} M.,  {Springel} V.,   {Grand} R.~J.~J.,  2018,
  \mn@doi [\mnras] {10.1093/mnras/sty1692}, \href
  {http://adsabs.harvard.edu/abs/2018MNRAS.479.3381B} {479, 3381}

\bibitem[\protect\citeauthoryear{{Chakrabarti} \& {Blitz}}{{Chakrabarti} \&
  {Blitz}}{2009}]{ChakrabartiandBlitz2009}
{Chakrabarti} S.,  {Blitz} L.,  2009, \mn@doi [\mnras]
  {10.1111/j.1745-3933.2009.00735.x}, \href
  {http://adsabs.harvard.edu/abs/2009MNRAS.399L.118C} {399, L118}

\bibitem[\protect\citeauthoryear{Chakrabarti, Bigiel, Chang  \&
  Blitz}{Chakrabarti et~al.}{2011}]{Chakrabarti11}
Chakrabarti S.,  Bigiel F.,  Chang P.,   Blitz L.,  2011, ApJ, 743, 35

\bibitem[\protect\citeauthoryear{Chang}{Chang}{2008}]{Chang2008}
Chang P.,  2008, ApJ, 684, 236

\bibitem[\protect\citeauthoryear{{Chang} \& {Chakrabarti}}{{Chang} \&
  {Chakrabarti}}{2011}]{ChangandChakrabarti2011}
{Chang} P.,  {Chakrabarti} S.,  2011, \mn@doi [\mnras]
  {10.1111/j.1365-2966.2011.19071.x}, \href
  {http://adsabs.harvard.edu/abs/2011MNRAS.416..618C} {416, 618}

\bibitem[\protect\citeauthoryear{{Cox}, {Primack}, {Jonsson}  \&
  {Somerville}}{{Cox} et~al.}{2004}]{Coxetal2004}
{Cox} T.~J.,  {Primack} J.,  {Jonsson} P.,   {Somerville} R.~S.,  2004, \mn@doi
  [\apjl] {10.1086/421905}, \href
  {http://adsabs.harvard.edu/abs/2004ApJ...607L..87C} {607, L87}

\bibitem[\protect\citeauthoryear{{Cox}, {Jonsson}, {Somerville}, {Primack}  \&
  {Dekel}}{{Cox} et~al.}{2008}]{Coxetal2008}
{Cox} T.~J.,  {Jonsson} P.,  {Somerville} R.~S.,  {Primack} J.~R.,   {Dekel}
  A.,  2008, \mn@doi [\mnras] {10.1111/j.1365-2966.2007.12730.x}, \href
  {http://adsabs.harvard.edu/abs/2008MNRAS.384..386C} {384, 386}

\bibitem[\protect\citeauthoryear{{D'Onghia}, {Madau}, {Vera-Ciro}, {Quillen}
  \& {Hernquist}}{{D'Onghia} et~al.}{2016}]{DOnghiaetal2016}
{D'Onghia} E.,  {Madau} P.,  {Vera-Ciro} C.,  {Quillen} A.,   {Hernquist} L.,
  2016, \mn@doi [\apj] {10.3847/0004-637X/823/1/4}, \href
  {https://ui.adsabs.harvard.edu/abs/2016ApJ...823....4D} {823, 4}

\bibitem[\protect\citeauthoryear{{Di Matteo}, {Bournaud}, {Martig}, {Combes},
  {Melchior}  \& {Semelin}}{{Di Matteo} et~al.}{2008}]{DiMatteoetal2008}
{Di Matteo} P.,  {Bournaud} F.,  {Martig} M.,  {Combes} F.,  {Melchior} A.~L.,
   {Semelin} B.,  2008, \mn@doi [\aap] {10.1051/0004-6361:200809480}, \href
  {https://ui.adsabs.harvard.edu/abs/2008A&A...492...31D} {492, 31}

\bibitem[\protect\citeauthoryear{Dobbs, Theis, Pringle  \& Bate}{Dobbs
  et~al.}{2010}]{Dobbsetal2010}
Dobbs C.~L.,  Theis C.,  Pringle J.~E.,   Bate M.~R.,  2010, MNRAS, 403, 625

\bibitem[\protect\citeauthoryear{{Edman}, {Barton}  \& {Bullock}}{{Edman}
  et~al.}{2012}]{Edmanetal2012}
{Edman} J.~P.,  {Barton} E.~J.,   {Bullock} J.~S.,  2012, \mn@doi [\mnras]
  {10.1111/j.1365-2966.2012.21335.x}, \href
  {https://ui.adsabs.harvard.edu/abs/2012MNRAS.424.1454E} {424, 1454}

\bibitem[\protect\citeauthoryear{{Gabor}, {Capelo}, {Volonteri}, {Bournaud},
  {Bellovary}, {Governato}  \& {Quinn}}{{Gabor} et~al.}{2016}]{Gaboretal2016}
{Gabor} J.~M.,  {Capelo} P.~R.,  {Volonteri} M.,  {Bournaud} F.,  {Bellovary}
  J.,  {Governato} F.,   {Quinn} T.,  2016, \mn@doi [\aap]
  {10.1051/0004-6361/201527143}, \href
  {https://ui.adsabs.harvard.edu/abs/2016A&A...592A..62G} {592, A62}

\bibitem[\protect\citeauthoryear{{Galloway} et~al.,}{{Galloway}
  et~al.}{2015}]{Gallowayetal2015}
{Galloway} M.~A.,  et~al., 2015, MNRAS, 448, 3442

\bibitem[\protect\citeauthoryear{Gingold \& Monaghan}{Gingold \&
  Monaghan}{1977}]{GingoldMonaghan77}
Gingold R.~A.,  Monaghan J.~J.,  1977, MNRAS, 181, 375

\bibitem[\protect\citeauthoryear{{Gonz{\'a}lez}, {Lares}, {Lambas}  \&
  {Valotto}}{{Gonz{\'a}lez} et~al.}{2006}]{Gonzalezetal2006}
{Gonz{\'a}lez} R.~E.,  {Lares} M.,  {Lambas} D.~G.,   {Valotto} C.,  2006,
  \mn@doi [\aap] {10.1051/0004-6361:20053277}, 445, 51

\bibitem[\protect\citeauthoryear{{Goulding} et~al.,}{{Goulding}
  et~al.}{2018}]{Gouldingetal2018}
{Goulding} A.~D.,  et~al., 2018, \mn@doi [PASJ] {10.1093/pasj/psx135}, \href
  {https://ui.adsabs.harvard.edu/abs/2018PASJ...70S..37G} {70, S37}

\bibitem[\protect\citeauthoryear{Haan, Schinnerer, Emsellem, García-Burrillo,
  Combes, Mundell  \& Rix}{Haan et~al.}{2009}]{Haanetal09}
Haan S.,  Schinnerer E.,  Emsellem E.,  García-Burrillo S.,  Combes F.,
  Mundell C.~G.,   Rix H.~W.,  2009, ApJ, 692, 1623

\bibitem[\protect\citeauthoryear{Hernquist}{Hernquist}{1990}]{Hernquist90}
Hernquist L.,  1990, ApJ, 356, 359

\bibitem[\protect\citeauthoryear{Hernquist \& Katz}{Hernquist \&
  Katz}{1985}]{HernquistKatz1989}
Hernquist L.,  Katz N.,  1985, ApJS, 149, 135

\bibitem[\protect\citeauthoryear{Hernquist \& Mihos}{Hernquist \&
  Mihos}{1995}]{HernquistMihos95}
Hernquist L.,  Mihos J.~C.,  1995, ApJ, 448, 41

\bibitem[\protect\citeauthoryear{{Holmberg}}{{Holmberg}}{1941}]{Holmberg1941}
{Holmberg} E.,  1941, \mn@doi [\apj] {10.1086/144344}, \href
  {http://adsabs.harvard.edu/abs/1941ApJ....94..385H} {94, 385}

\bibitem[\protect\citeauthoryear{Hopkins \& Quataert}{Hopkins \&
  Quataert}{2010}]{HopkinsQuataert2010}
Hopkins P.~F.,  Quataert E.,  2010, MNRAS, 407, 1529

\bibitem[\protect\citeauthoryear{{Hopkins}, {Kere{\v s}}, {Ma}  \&
  {Quataert}}{{Hopkins} et~al.}{2010}]{Hopkinsetal2010}
{Hopkins} P.~F.,  {Kere{\v s}} D.,  {Ma} C.-P.,   {Quataert} E.,  2010, \mn@doi
  [\mnras] {10.1111/j.1365-2966.2009.15700.x}, \href
  {http://adsabs.harvard.edu/abs/2010MNRAS.401.1131H} {401, 1131}

\bibitem[\protect\citeauthoryear{{Hu} \& {Sijacki}}{{Hu} \&
  {Sijacki}}{2018}]{HuandSijacki2018}
{Hu} S.,  {Sijacki} D.,  2018, \mn@doi [\mnras] {10.1093/mnras/sty1183}, \href
  {https://ui.adsabs.harvard.edu/abs/2018MNRAS.478.1576H} {478, 1576}

\bibitem[\protect\citeauthoryear{Jog \& Combes}{Jog \&
  Combes}{2009}]{JogCombes09}
Jog C.~J.,  Combes F.,  2009, Physics Reports, 471, 75

\bibitem[\protect\citeauthoryear{{Kannan}, {Macci{\`o}}, {Pasquali}, {Moster}
  \& {Walter}}{{Kannan} et~al.}{2012}]{Kannanetal2012}
{Kannan} R.,  {Macci{\`o}} A.~V.,  {Pasquali} A.,  {Moster} B.~P.,   {Walter}
  F.,  2012, \mn@doi [\apj] {10.1088/0004-637X/746/1/10}, \href
  {https://ui.adsabs.harvard.edu/abs/2012ApJ...746...10K} {746, 10}

\bibitem[\protect\citeauthoryear{Karachentsev}{Karachentsev}{1972}]{Karachentsev1972}
Karachentsev I.~D.,  1972, Soobshch. Spets. Astrofiz. Obs., Vyp. 7, 92

\bibitem[\protect\citeauthoryear{{Kaviraj}}{{Kaviraj}}{2014}]{Kaviraj2014}
{Kaviraj} S.,  2014, \mn@doi [\mnras] {10.1093/mnrasl/slt136}, \href
  {http://adsabs.harvard.edu/abs/2014MNRAS.437L..41K} {437, L41}

\bibitem[\protect\citeauthoryear{Kazantzidis, Bullock, Zentner, Kravtsov  \&
  Moustakas}{Kazantzidis et~al.}{2008}]{Kazantzidis08}
Kazantzidis S.,  Bullock J.~S.,  Zentner A.~R.,  Kravtsov A.~V.,   Moustakas
  L.~A.,  2008, ApJ, 688, 254

\bibitem[\protect\citeauthoryear{Kazantzidis, Zentner, Kravtsov, Bullock  \&
  Debattista}{Kazantzidis et~al.}{2009}]{Kazantzidis09}
Kazantzidis S.,  Zentner A.~R.,  Kravtsov A.~V.,  Bullock J.~S.,   Debattista
  V.~P.,  2009, ApJ, 700, 1896

\bibitem[\protect\citeauthoryear{{Khochfar} \& {Burkert}}{{Khochfar} \&
  {Burkert}}{2006}]{KhochfarBurkert2006}
{Khochfar} S.,  {Burkert} A.,  2006, A\&A, 445, 603

\bibitem[\protect\citeauthoryear{Kim \& Kim}{Kim \& Kim}{2014}]{KimKim2014}
Kim Y.,  Kim W.-T.,  2014, MNRAS, 440, 208

\bibitem[\protect\citeauthoryear{{Kim}, {Seo}  \& {Kim}}{{Kim}
  et~al.}{2012}]{KimSeoKim2012}
{Kim} W.~T.,  {Seo} W.~Y.,   {Kim} Y.,  2012, ApJ, 758, 14

\bibitem[\protect\citeauthoryear{Klypin, Kravtsov, Valenzuela  \& Prada}{Klypin
  et~al.}{1999}]{Klypinetal1999}
Klypin A.,  Kravtsov A.~V.,  Valenzuela O.,   Prada F.,  1999, ApJ, 522, 82

\bibitem[\protect\citeauthoryear{Klypin, Zhao  \& Somerville}{Klypin
  et~al.}{2002}]{KlypinZhaoSomerville02}
Klypin A.,  Zhao H.,   Somerville R.,  2002, ApJ, 573, 597

\bibitem[\protect\citeauthoryear{{Koda}, {Milosavljevi{\'c}}  \&
  {Shapiro}}{{Koda} et~al.}{2009}]{Kodaetal2009}
{Koda} J.,  {Milosavljevi{\'c}} M.,   {Shapiro} P.~R.,  2009, \mn@doi [\apj]
  {10.1088/0004-637X/696/1/254}, \href
  {https://ui.adsabs.harvard.edu/abs/2009ApJ...696..254K} {696, 254}

\bibitem[\protect\citeauthoryear{{Kotarba}, {Lesch}, {Dolag}, {Naab},
  {Johansson}, {Donnert}  \& {Stasyszyn}}{{Kotarba}
  et~al.}{2011}]{Kotarbaetal2011}
{Kotarba} H.,  {Lesch} H.,  {Dolag} K.,  {Naab} T.,  {Johansson} P.~H.,
  {Donnert} J.,   {Stasyszyn} F.~A.,  2011, \mn@doi [\mnras]
  {10.1111/j.1365-2966.2011.18932.x}, \href
  {http://adsabs.harvard.edu/abs/2011MNRAS.415.3189K} {415, 3189}

\bibitem[\protect\citeauthoryear{{Kyziropoulos}, {Efthymiopoulos}, {Gravvanis}
  \& {Patsis}}{{Kyziropoulos} et~al.}{2016}]{Kyziropoulosetal2016}
{Kyziropoulos} P.~E.,  {Efthymiopoulos} C.,  {Gravvanis} G.~A.,   {Patsis}
  P.~A.,  2016, \mn@doi [\mnras] {10.1093/mnras/stw2084}, \href
  {https://ui.adsabs.harvard.edu/abs/2016MNRAS.463.2210K} {463, 2210}

\bibitem[\protect\citeauthoryear{{Lipnicky}, {Chakrabarti}  \&
  {Chang}}{{Lipnicky} et~al.}{2018}]{Lipnickyetal2018}
{Lipnicky} A.,  {Chakrabarti} S.,   {Chang} P.,  2018, \mn@doi [\mnras]
  {10.1093/mnras/sty2330}, \href
  {http://adsabs.harvard.edu/abs/2018MNRAS.481.2590L} {481, 2590}

\bibitem[\protect\citeauthoryear{{Lora}, {Just}, {S{\'a}nchez-Salcedo}  \&
  {Grebel}}{{Lora} et~al.}{2012}]{Loraetal2012}
{Lora} V.,  {Just} A.,  {S{\'a}nchez-Salcedo} F.~J.,   {Grebel} E.~K.,  2012,
  \mn@doi [\apj] {10.1088/0004-637X/757/1/87}, \href
  {https://ui.adsabs.harvard.edu/abs/2012ApJ...757...87L} {757, 87}

\bibitem[\protect\citeauthoryear{{Mapelli}, {Rampazzo}  \& {Marino}}{{Mapelli}
  et~al.}{2015}]{Mapellietal2015}
{Mapelli} M.,  {Rampazzo} R.,   {Marino} A.,  2015, \mn@doi [\aap]
  {10.1051/0004-6361/201425315}, 575, A16

\bibitem[\protect\citeauthoryear{{Matthews}, {van Driel}  \&
  {Gallagher}}{{Matthews} et~al.}{1998}]{Matthewsetal1998}
{Matthews} L.~D.,  {van Driel} W.,   {Gallagher} J.~S. I.,  1998, \mn@doi [\aj]
  {10.1086/300492}, \href
  {https://ui.adsabs.harvard.edu/abs/1998AJ....116.1169M} {116, 1169}

\bibitem[\protect\citeauthoryear{McMillan \& Dehnen}{McMillan \&
  Dehnen}{2007}]{McMillanDehnen07}
McMillan P.~J.,  Dehnen W.,  2007, MNRAS, 378, 541

\bibitem[\protect\citeauthoryear{Mihos \& Hernquist}{Mihos \&
  Hernquist}{1994}]{MihosHernquist94}
Mihos J.~C.,  Hernquist L.,  1994, ApJ, 425, L13

\bibitem[\protect\citeauthoryear{Mo, van~den Bosch  \& White}{Mo
  et~al.}{2010}]{MoVanDenBoschWhite2010}
Mo H.,  van~den Bosch F.,   White S. D.~M.,  2010, {Galaxy Formation and
  Evolution}.
Cambridge University Press

\bibitem[\protect\citeauthoryear{{Moetazedian} \& {Just}}{{Moetazedian} \&
  {Just}}{2016}]{MoetazedianandJust2016}
{Moetazedian} R.,  {Just} A.,  2016, \mn@doi [\mnras] {10.1093/mnras/stw764},
  \href {https://ui.adsabs.harvard.edu/abs/2016MNRAS.459.2905M} {459, 2905}

\bibitem[\protect\citeauthoryear{Montuori, Matteo, Lehnert, Combes  \&
  Semelin}{Montuori et~al.}{2010}]{Montuori2010}
Montuori M.,  Matteo P.~D.,  Lehnert M.~D.,  Combes F.,   Semelin B.,  2010,
  A\&A, 518, A56

\bibitem[\protect\citeauthoryear{Moore, Ghigna, Governato, Lake, Quinn, Stadel
  \& Tozzi}{Moore et~al.}{1999}]{Mooreetal1999}
Moore B.,  Ghigna S.,  Governato F.,  Lake G.,  Quinn T.,  Stadel J.,   Tozzi
  P.,  1999, ApJ, 524, L19

\bibitem[\protect\citeauthoryear{Navarro, Frenk  \& White}{Navarro
  et~al.}{1996}]{NFW96}
Navarro J.~F.,  Frenk C.,   White S. D.~M.,  1996, ApJ, 462, 462

\bibitem[\protect\citeauthoryear{{Pettitt} \& {Wadsley}}{{Pettitt} \&
  {Wadsley}}{2018}]{Pettittetal2018}
{Pettitt} A.~R.,  {Wadsley} J.~W.,  2018, \mn@doi [\mnras]
  {10.1093/mnras/stx3129}, \href
  {http://adsabs.harvard.edu/abs/2018MNRAS.474.5645P} {474, 5645}

\bibitem[\protect\citeauthoryear{{Pettitt}, {Tasker}  \& {Wadsley}}{{Pettitt}
  et~al.}{2016}]{Pettittetal2016}
{Pettitt} A.~R.,  {Tasker} E.~J.,   {Wadsley} J.~W.,  2016, \mn@doi [\mnras]
  {10.1093/mnras/stw588}, \href
  {http://adsabs.harvard.edu/abs/2016MNRAS.458.3990P} {458, 3990}

\bibitem[\protect\citeauthoryear{{Price}}{{Price}}{2012}]{Price2012}
{Price} D.~J.,  2012, \mn@doi [Journal of Computational Physics]
  {10.1016/j.jcp.2010.12.011}, \href
  {http://adsabs.harvard.edu/abs/2012JCoPh.231..759P} {231, 759}

\bibitem[\protect\citeauthoryear{{Purcell}, {Bullock}  \&
  {Kazantzidis}}{{Purcell} et~al.}{2010}]{Purcelletal2010}
{Purcell} C.~W.,  {Bullock} J.~S.,   {Kazantzidis} S.,  2010, MNRAS, 404, 1711

\bibitem[\protect\citeauthoryear{Purcell, Bullock, Tollerud, Rocha  \&
  Chakrabarti}{Purcell et~al.}{2011}]{Purcelletal2011}
Purcell C.~W.,  Bullock J.~S.,  Tollerud E.~J.,  Rocha M.,   Chakrabarti S.,
  2011, Nature, 477, 301

\bibitem[\protect\citeauthoryear{{Qu}, {Di Matteo}, {Lehnert}  \& {van
  Driel}}{{Qu} et~al.}{2011}]{Quetal2011}
{Qu} Y.,  {Di Matteo} P.,  {Lehnert} M.~D.,   {van Driel} W.,  2011, MNRAS,
  740, 101

\bibitem[\protect\citeauthoryear{R.~C.~Kennicutt}{R.~C.~Kennicutt}{1998}]{Kennicutt98}
R.~C.~Kennicutt J.,  1998, ARA\&A, 36, 189

\bibitem[\protect\citeauthoryear{{Ram{\'o}n-Fox}}{{Ram{\'o}n-Fox}}{2019}]{Ramon-Fox2019}
{Ram{\'o}n-Fox} F.~G.,  2019, PhD thesis, University of St Andrews

\bibitem[\protect\citeauthoryear{{Regan} \& {Teuben}}{{Regan} \&
  {Teuben}}{2004}]{ReganTeuben2004}
{Regan} M.~W.,  {Teuben} P.~J.,  2004, ApJ, 600, 595

\bibitem[\protect\citeauthoryear{{Renaud}, {Bournaud}  \& {Duc}}{{Renaud}
  et~al.}{2015}]{Renaudetal2015}
{Renaud} F.,  {Bournaud} F.,   {Duc} P.-A.,  2015, \mn@doi [\mnras]
  {10.1093/mnras/stu2208}, \href
  {http://adsabs.harvard.edu/abs/2015MNRAS.446.2038R} {446, 2038}

\bibitem[\protect\citeauthoryear{Richter \& Sancisi}{Richter \&
  Sancisi}{1994}]{RichterSancisi94}
Richter O.~G.,  Sancisi R.,  1994, A\&A, 290L, L9

\bibitem[\protect\citeauthoryear{Rix \& Zaritsky}{Rix \&
  Zaritsky}{1995}]{RixZaritsky95}
Rix H.-W.,  Zaritsky D.,  1995, ApJ, 447, 82

\bibitem[\protect\citeauthoryear{{Rudnick} \& {Rix}}{{Rudnick} \&
  {Rix}}{1998}]{RudnickandRix1998}
{Rudnick} G.,  {Rix} H.-W.,  1998, \mn@doi [\aj] {10.1086/300518}, \href
  {https://ui.adsabs.harvard.edu/abs/1998AJ....116.1163R} {116, 1163}

\bibitem[\protect\citeauthoryear{{Rudnick}, {Rix}  \& {Kennicutt}}{{Rudnick}
  et~al.}{2000}]{Rudnicketal2000}
{Rudnick} G.,  {Rix} H.-W.,   {Kennicutt} Robert~C. J.,  2000, \mn@doi [\apj]
  {10.1086/309169}, \href
  {https://ui.adsabs.harvard.edu/abs/2000ApJ...538..569R} {538, 569}

\bibitem[\protect\citeauthoryear{{Satyapal}, {Ellison}, {McAlpine}, {Hickox},
  {Patton}  \& {Mendel}}{{Satyapal} et~al.}{2014}]{Satyapaletal2014}
{Satyapal} S.,  {Ellison} S.~L.,  {McAlpine} W.,  {Hickox} R.~C.,  {Patton}
  D.~R.,   {Mendel} J.~T.,  2014, \mn@doi [\mnras] {10.1093/mnras/stu650},
  \href {https://ui.adsabs.harvard.edu/abs/2014MNRAS.441.1297S} {441, 1297}

\bibitem[\protect\citeauthoryear{{Sawala}, {Pihajoki}, {Johansson}, {Frenk},
  {Navarro}, {Oman}  \& {White}}{{Sawala} et~al.}{2017}]{Sawalaetal2017}
{Sawala} T.,  {Pihajoki} P.,  {Johansson} P.~H.,  {Frenk} C.~S.,  {Navarro}
  J.~F.,  {Oman} K.~A.,   {White} S. D.~M.,  2017, \mn@doi [\mnras]
  {10.1093/mnras/stx360}, \href
  {https://ui.adsabs.harvard.edu/abs/2017MNRAS.467.4383S} {467, 4383}

\bibitem[\protect\citeauthoryear{{Schaye} \& {Dalla Vecchia}}{{Schaye} \&
  {Dalla Vecchia}}{2008}]{SchayeandDallaVecchia2008}
{Schaye} J.,  {Dalla Vecchia} C.,  2008, \mn@doi [\mnras]
  {10.1111/j.1365-2966.2007.12639.x}, \href
  {https://ui.adsabs.harvard.edu/abs/2008MNRAS.383.1210S} {383, 1210}

\bibitem[\protect\citeauthoryear{{Shah}, {Bekki}, {Vinsen}  \& {Foster}}{{Shah}
  et~al.}{2019}]{Shahetal2019}
{Shah} M.,  {Bekki} K.,  {Vinsen} K.,   {Foster} S.,  2019, \mn@doi [\mnras]
  {10.1093/mnras/sty2897}, \href
  {http://adsabs.harvard.edu/abs/2019MNRAS.482.4188S} {482, 4188}

\bibitem[\protect\citeauthoryear{Shlosman, Begelman  \& Frank}{Shlosman
  et~al.}{1990}]{Shlosmanetal90}
Shlosman I.,  Begelman M.~C.,   Frank J.,  1990, Nature, 345, 679

\bibitem[\protect\citeauthoryear{{Smilgys} \& {Bonnell}}{{Smilgys} \&
  {Bonnell}}{2017}]{SmilgysandBonnell2017}
{Smilgys} R.,  {Bonnell} I.~A.,  2017, \mn@doi [\mnras]
  {10.1093/mnras/stx2396}, \href
  {https://ui.adsabs.harvard.edu/abs/2017MNRAS.472.4982S} {472, 4982}

\bibitem[\protect\citeauthoryear{{Springel}}{{Springel}}{2000}]{Springel2000}
{Springel} V.,  2000, \mn@doi [\mnras] {10.1046/j.1365-8711.2000.03187.x},
  \href {http://adsabs.harvard.edu/abs/2000MNRAS.312..859S} {312, 859}

\bibitem[\protect\citeauthoryear{Springel}{Springel}{2005}]{Springel05}
Springel V.,  2005, MNRAS, 364, 1105

\bibitem[\protect\citeauthoryear{{Springel} \& {Hernquist}}{{Springel} \&
  {Hernquist}}{2005}]{SpringelandHernquist2005}
{Springel} V.,  {Hernquist} L.,  2005, \mn@doi [\apjl] {10.1086/429486}, \href
  {http://adsabs.harvard.edu/abs/2005ApJ...622L...9S} {622, L9}

\bibitem[\protect\citeauthoryear{{Starkenburg}, {Helmi}  \&
  {Sales}}{{Starkenburg} et~al.}{2016}]{Starkenburgetal2016}
{Starkenburg} T.~K.,  {Helmi} A.,   {Sales} L.~V.,  2016, \mn@doi [\aap]
  {10.1051/0004-6361/201528066}, \href
  {https://ui.adsabs.harvard.edu/abs/2016A&A...595A..56S} {595, A56}

\bibitem[\protect\citeauthoryear{{Taniguchi}}{{Taniguchi}}{1999}]{Taniguchi1999}
{Taniguchi} Y.,  1999, \mn@doi [\apj] {10.1086/307814}, \href
  {https://ui.adsabs.harvard.edu/abs/1999ApJ...524...65T} {524, 65}

\bibitem[\protect\citeauthoryear{Teuben}{Teuben}{1995}]{Teuben95}
Teuben P.~J.,  1995, in Astronomical Data Analysis Software and Systems IV. 77.
PASP Conference Series, p.~398

\bibitem[\protect\citeauthoryear{{Toomre} \& {Toomre}}{{Toomre} \&
  {Toomre}}{1972}]{ToomreandToomre1972}
{Toomre} A.,  {Toomre} J.,  1972, \mn@doi [\apj] {10.1086/151823}, \href
  {http://adsabs.harvard.edu/abs/1972ApJ...178..623T} {178, 623}

\bibitem[\protect\citeauthoryear{{Torrey}, {Cox}, {Kewley}  \&
  {Hernquist}}{{Torrey} et~al.}{2012}]{Torreyetal2012}
{Torrey} P.,  {Cox} T.~J.,  {Kewley} L.,   {Hernquist} L.,  2012, \mn@doi
  [\apj] {10.1088/0004-637X/746/1/108}, \href
  {http://adsabs.harvard.edu/abs/2012ApJ...746..108T} {746, 108}

\bibitem[\protect\citeauthoryear{{Toth} \& {Ostriker}}{{Toth} \&
  {Ostriker}}{1992}]{TothandOstriker1992}
{Toth} G.,  {Ostriker} J.~P.,  1992, \mn@doi [\apj] {10.1086/171185}, \href
  {https://ui.adsabs.harvard.edu/abs/1992ApJ...389....5T} {389, 5}

\bibitem[\protect\citeauthoryear{Velazquez \& White}{Velazquez \&
  White}{1999}]{VelazquezWhite1999}
Velazquez H.,  White S. D.~M.,  1999, MNRAS, 304, 254

\bibitem[\protect\citeauthoryear{{Wagner-Kaiser}, {De Maio}, {Sarajedini}  \&
  {Chakrabarti}}{{Wagner-Kaiser} et~al.}{2014}]{Wagner-Kaiseretal2014}
{Wagner-Kaiser} R.,  {De Maio} T.,  {Sarajedini} A.,   {Chakrabarti} S.,  2014,
  \mn@doi [\mnras] {10.1093/mnras/stu1327}, \href
  {https://ui.adsabs.harvard.edu/abs/2014MNRAS.443.3260W} {443, 3260}

\bibitem[\protect\citeauthoryear{{Wang} \& {White}}{{Wang} \&
  {White}}{2012}]{WangandWhite2012}
{Wang} W.,  {White} S.~D.~M.,  2012, \mn@doi [\mnras]
  {10.1111/j.1365-2966.2012.21256.x}, \href
  {http://adsabs.harvard.edu/abs/2012MNRAS.424.2574W} {424, 2574}

\bibitem[\protect\citeauthoryear{Wang, Klessen, Dullemond, van~den Bosch  \&
  Fuchs}{Wang et~al.}{2010}]{Wang10}
Wang H.,  Klessen R.~S.,  Dullemond C.~P.,  van~den Bosch F.~C.,   Fuchs B.,
  2010, MNRAS, 407, 705

\bibitem[\protect\citeauthoryear{{Wetzel}}{{Wetzel}}{2011}]{Wetzel2011}
{Wetzel} A.~R.,  2011, MNRAS, 412, 49

\bibitem[\protect\citeauthoryear{{Willett} et~al.,}{{Willett}
  et~al.}{2015}]{Willettetal2015}
{Willett} K.~W.,  et~al., 2015, \mn@doi [\mnras] {10.1093/mnras/stv307}, \href
  {https://ui.adsabs.harvard.edu/abs/2015MNRAS.449..820W} {449, 820}

\bibitem[\protect\citeauthoryear{Zaritsky \& Rix}{Zaritsky \&
  Rix}{1997}]{ZaritskyRix97}
Zaritsky D.,  Rix H.~W.,  1997, ApJ, 477, 118

\bibitem[\protect\citeauthoryear{{Zaritsky} et~al.,}{{Zaritsky}
  et~al.}{2013}]{Zaritskyetal2013}
{Zaritsky} D.,  et~al., 2013, ApJ, 772, 135

\bibitem[\protect\citeauthoryear{{Zentner}, {Kravtsov}, {Gnedin}  \&
  {Klypin}}{{Zentner} et~al.}{2005}]{Zentneretal2005}
{Zentner} A.~R.,  {Kravtsov} Z.~V.,  {Gnedin} O.~Y.,   {Klypin} A.~A.,  2005,
  ApJ, 629, 219

\makeatother
\end{thebibliography}
%%%%%%%%%%%%%%%%%%%%%%%%%%%%%

\appendix

%%%%%%%%%%%%%%%%%%%%%%%%%%%%%%%%%%%%%%%%%%%%%%%%%%
%%%%%%%%%%%%%%%%%%%%%%%%%%%%%%%%%%%%%%%%%%%%%%%%%%
%%%%%%%%%%%%%%%%%%%%%%%%%%%%%%%%%%%%%%%%%%%%%%%%%%
\section{Resolution Tests}
\label{sec:A-Res-Tests}

In order to test the effect of changing the gas resolution on our results, we repeated some of the simulations with $2$ and $4$ times higher resolution. In this section, we briefly describe the procedure and summarize the main results of our tests.

We selected the following cases: 1) the interaction of Model A and Satellite 2 in prograde Orbit 2, and 2) the one of Model A and Satellite 1 in retrograde Orbit 3. This choice was motivated by the fact that the first one produces the most significant change in the mass distribution, while the second one has a much lower the effect. This allows to compare to extreme results. The gas resolution of the simulations was increased by specifying a split factor $s$ such that the new number of gas particles is $N_{\mathrm{new}} = s N_{\mathrm{orig}}$, where $N_{\mathrm{orig}}$ is the original gas particles. Then, the new gas particle mass is $m_{\mathrm{new}} = m_{\mathrm{orig}}/s$. The new particle positions were specified by sampling the new particles using the kernel function of the original one. We tested simulations with $s = 2$ and $s = 4$. The non-gaseous components were not changed for these tests. A similar technique has been tested in \citet{SmilgysandBonnell2017}.

\begin{figure}
	\begin{center}
		\includegraphics[width=0.9\linewidth,trim=10mm 5mm 10mm 10mm,clip=true]{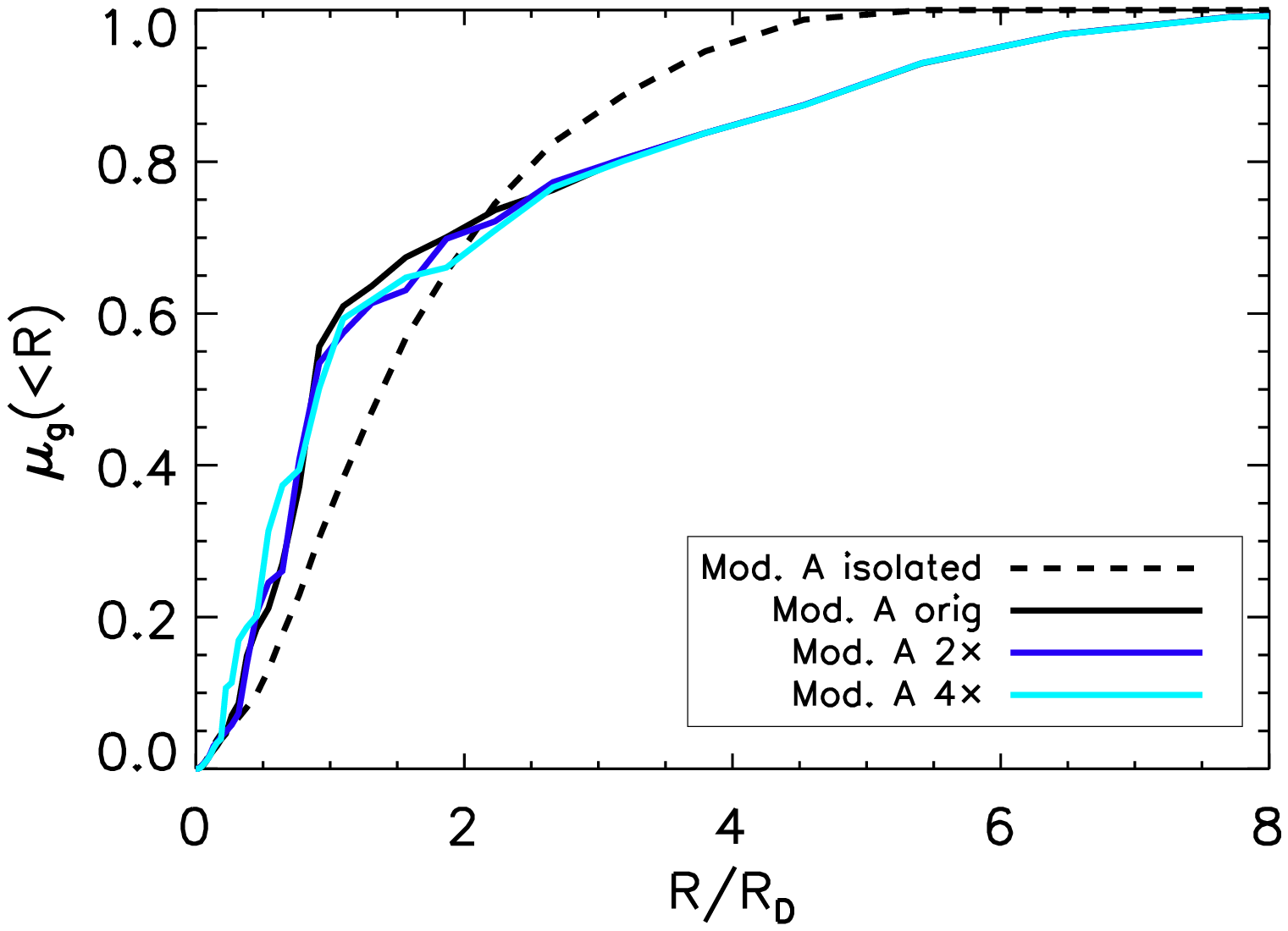}
		
		\includegraphics[width=0.9\linewidth,trim=10mm 5mm 10mm 10mm,clip=true]{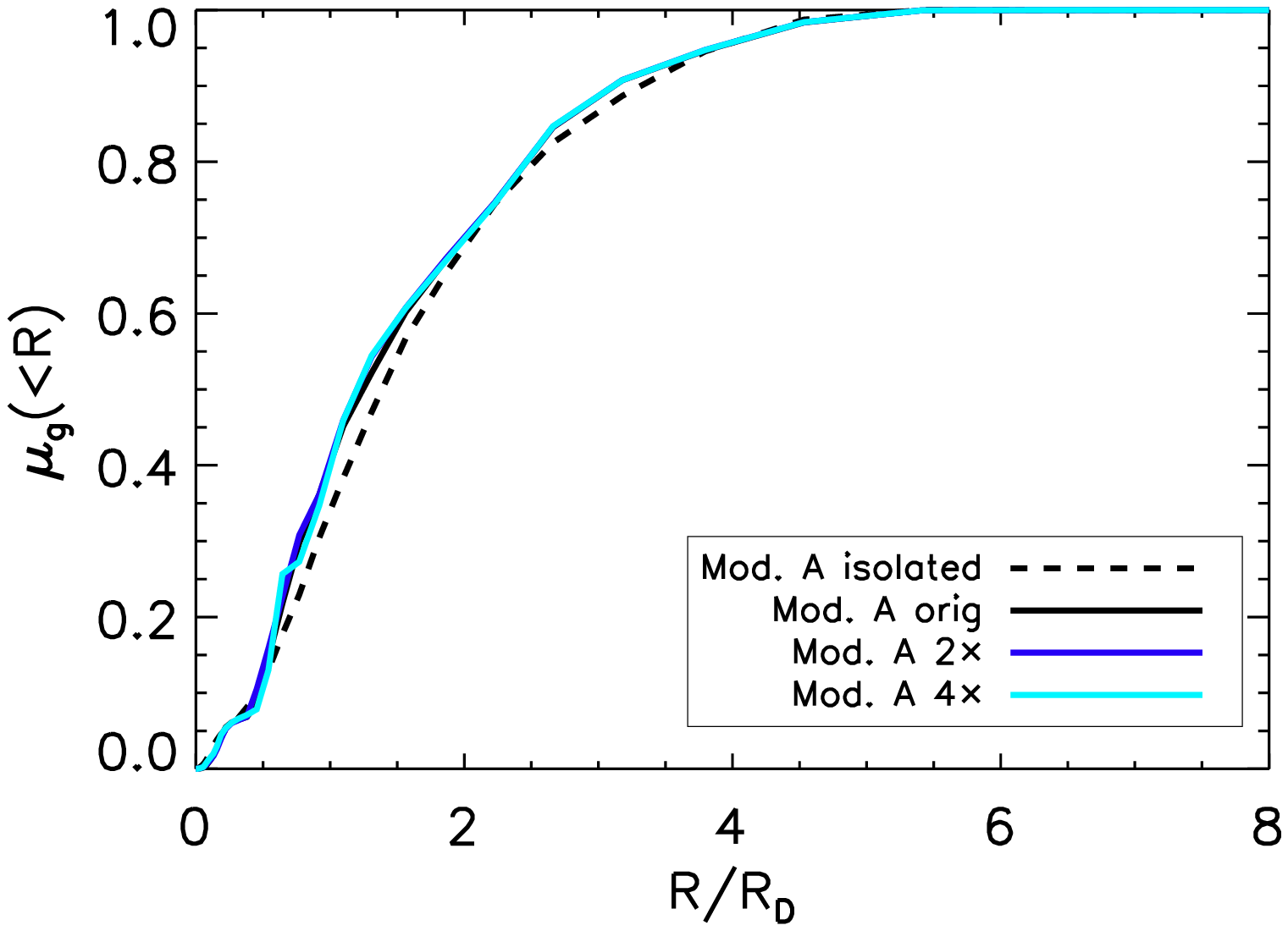}
	
		\caption{Integrated mass distributions $\mu_g$ for the resolution test simulations with Model A and Satellite 2 with prograde Orbit 2 at $t \approx 0.65\tau$ in the top panel and for Model A and Satellite 1 with retrograde Orbit 3 at $t \approx 0.45\tau$ in the bottom panel. The black dashed line is $\mu_g$ for the isolated model, the solid black one the one for the original simulation, the dark blue the one for the $2\times$ higher resolution simulation, and the light blue the one for the $4\times$ higher resolution simulation.}
		\label{fig:res-test-1}
	\end{center}
\end{figure}

In general, the higher-resolution simulations show the same large-scale morphological features and induced spiral arms. The top panel of Figure \ref{fig:res-test-1} shows the integrated mass function $\mu_g(<R)$ at $t \approx 0.65\tau$ for case 1) and the bottom panel of Figure \ref{fig:res-test-1} plots $\mu_g(<R)$ at $t \approx 0.45\tau$ for case 2). Both show essentially the same features regardless of the resolution used, although for case 1) there are some slight differences between each simulation. We note that, for computational constraints, we obtained $\mu_g(<R)$ at earlier times compared to those plotted in Figure \ref{fig:mass-distributions-model-A}. Some of the simulations develop some clumps particularly from gas in the spiral arms. After a qualitative inspection of the morphology of the higher resolution simulations, we note that the number clumps formed varies with the resolution, which is to be expected. We noticed that it can have some effect on the shape of the $\mu_g(<R)$. The position of these clumps relative to the galaxy's centre can introduce some step-like features in $\mu_g(<R)$, particularly if they have gathered a significant amount of gas. However, we emphasise that the large-scale features and flows do not change significantly with resolution according to our tests.

\bsp
\label{lastpage}
\end{document}